\documentclass[sigconf, anonymous=false]{acmart}

\usepackage{epsfig,endnotes}
\usepackage{type1cm}	
\usepackage{graphicx}   
\usepackage{xspace}     
\usepackage{balance}    
\usepackage{booktabs}   
\usepackage[tableposition=top,font={small}]{caption}  
\usepackage{siunitx}	
\usepackage{epsfig}

\usepackage{framed}
\usepackage{balance}
\usepackage{newfloat}
\usepackage{enumerate}
\usepackage[vlined,linesnumbered,ruled,noend]{algorithm2e}
\usepackage{lscape}
\usepackage{subfigure}
\usepackage{array}
\usepackage{pifont}
\usepackage{color, colortbl}
\usepackage{alltt}
\usepackage{listings}
\usepackage{balance}
\usepackage{multirow}
\usepackage{ifthen}
\usepackage{multirow}

\newcommand{\CL}{$CL$\xspace}

\newcommand{\descr}[1]{\smallskip\noindent\textbf{#1}}

\setcopyright{none}
\renewcommand\footnotetextcopyrightpermission[1]{}
\begin{document}

\title{%
  Modeling Aggression Propagation on Social Media \\
  \large Please cite: https://doi.org/10.1016/j.osnem.2021.100137}

\author{\LARGE Chrysoula Terizi\textsuperscript{1}, Despoina Chatzakou\textsuperscript{2}, Evaggelia Pitoura\textsuperscript{1}, Panayiotis Tsaparas\textsuperscript{1}, Nicolas Kourtellis\textsuperscript{3}}
\affiliation{
	\institution{\textsuperscript{1}University of Ioannina, Greece, \textsuperscript{2}{Centre for Research and Technology Hellas, Greece}, 
 \textsuperscript{3}{Telefonica Research, Spain}}
}
\email{chterizi@cs.uoi.gr, dchatzakou@iti.gr, pitoura@cs.uoi.gr, tsap@cs.uoi.gr, nicolas.kourtellis@telefonica.com}

\begin{abstract}
Cyberaggression has been studied in various contexts and online social platforms, and modeled on different data using state-of-the-art machine and deep learning algorithms to enable automatic detection and blocking of this behavior.
Users can be influenced to act aggressively or even bully others because of elevated toxicity and aggression in their own (online) social circle.
In effect, this behavior can propagate from one user and neighborhood to another, and therefore, spread in the network.
Interestingly, to our knowledge, no work has modeled the network dynamics of aggressive behavior.

In this paper, we take a first step towards this direction by studying propagation of aggression on social media using opinion dynamics.
We propose ways to model how aggression may propagate from one user to another, depending on how each user is connected to other aggressive or regular users.
Through extensive simulations on Twitter data, we study how aggressive behavior could propagate in the network.
We validate our models with crawled and annotated ground truth data, reaching up to 80\% $AUC$, and discuss the results and implications of our work.
\end{abstract}


\maketitle

\section{Introduction}

Online aggression has spiked in the last few years, with many reports of such behavior across different contexts~\cite{cyberbullying-facts,passive-aggression}.
Indeed, cyberaggression can potentially manifest in any type of platform, regardless of the target audience and utility or purpose envisioned for the platform.
In fact, such behavior has been observed in different online social media platforms such as Twitter~\cite{meanBirds,meanBirdsTweb2019}, Instagram~\cite{Hosseinmardi2015}, YouTube~\cite{Chen2012DetectingOffensiveLanguage}, Yahoo Finance~\cite{Djuric2015HateSpeechDetection}, Yahoo Answers~\cite{kayes2015ya-abuse}, 4chan~\cite{4chan}, and across various demographics (e.g., teenagers vs. adults, men vs. women, etc.~\cite{cyberbullying-facts,Smith2008CyberbullyingNature,Campbell1987ModelsOfAngerAndAggression}).
Interestingly, it is difficult to find a generally accepted definition of cyberaggression across disciplines.
As argued in~\cite{corcoran2015cyberbullying}, there are different ways to define online aggression, depending on frequency or severity of the behavior, power difference of the victim with aggressor, etc.

As shown in~\cite{Henneberger2017}, users can be influenced to act aggressively and even bully others because of elevated toxicity and aggression in their own social circle.
This behavior can manifest in a similar fashion in the online world as well, and aggression can propagate from one user and neighborhood to another, thus, spreading in the network.
In fact, some early studies in sociology and psychology already proposed models of computer abuse based on theories of \textit{social learning}, \textit{social bonds}, and \textit{planned behavior}~\cite{lee2002holistic}.
To this end, some early work by~\cite{squicciarini2015bully-pairwise} has studied the pairwise interactions between users and how bullies can influence others, in an online social network.
However, to our knowledge, no work has modeled the network dynamics of aggressive behavior using formal propagation or diffusion models, and study how users' online neighbors and whole social circles affect the propagation of aggression through the network in time.
This paper takes the first, but crucial, steps to investigate pending fundamental questions such as:
\emph{How can aggressive behavior propagate from one user or neighborhood in the network to another?}
\emph{What propagation model and parameters could best represent the aggression diffusion and its intensity?}

There are several formal models in literature to study propagation of information through a network, such as opinion dynamics~\cite{2016arXiv160506326S}, Linear Threshold (LT) and Independent Cascades (IC)~\cite{kempe2003maximizing,srivastava2015social}.
Given their suitability for behavior modeling, in this paper, we focus on opinion dynamics, widely used to model how opinions propagate through a network, and leave $LT$ and $IC$ models for future work.
We propose new methods for modeling aggression propagation based on classical opinion models, and accommodate how aggression may propagate from one user to another.
Indeed, aggression can manifest in a network in various ways: one-time harassment by single (or multiple) contact(s), repetitive bullying behavior by single (or multiple) contact(s), etc.
To make the problem tractable, we study aggression propagation in its minimal form of one-time aggressive behavior exhibited from one user to another.
If this behavior repeats between the same users, the aggression effect is stronger and cumulative, i.e., bullying~\cite{meanBirds,meanBirdsTweb2019}.
Thus, we opt for models that consider important factors such as how much the second user is exposed to aggressive behavior of the first user or its neighborhood, popularity of interacting users, internal user state, etc.
We validate the models' performance on real Twitter data to measure their ability to model this behavior through time.
Our contributions are as follows:
\begin{itemize}
\item We formally present the problem of aggression propagation in a social network, and the necessary assumptions to study it in an algorithmic fashion on a network of users.
\item We propose new network algorithms to model aggression propagation based on opinion dynamics methods informed by properties of aggression found in literature on psychology, sociology and computational social science.
\item We implement these methods into a framework that simulates aggression propagation in a network, while controlling for various experimental factors such as:
1) social network used, 
2) propagation model applied, 
3) selection and ordering of users or edges affected by the propagation.
This framework can be applied in different social networks, given appropriate data to bootstrap the models.
\item We present extensive experimentation with the simulating framework and Twitter data, and show how model performance depends on the various factors controlled.
We find that methods which consider direct interactions between users and users' internal aggression state, better model aggression and how it could evolve on Twitter.
We validate the top-performing methods with a case study on Twitter data and measure up to AUC=80\% on predicting which users will be aggressive or not in the future.
We discuss implications of our findings for curbing cyberaggression on Twitter and other networks with similar structure.
\item We open-source our simulation framework and data for reproducibility and extensibility purposes\footnote{github.com/cterizi/Modeling-Aggression-Propagation-on-Social-Media}.
\end{itemize}

\section{Modeling Aggression Propagation}\label{sec:models}

Aggression has been well studied in the past, in online and offline contexts, from sociologists and psychologists~\cite{Smith2008CyberbullyingNature,corcoran2015cyberbullying,FARVER1996333,Bandura1961,EmotionalBehavioralDisorders,Pieschl2013241,GAM} and computational social or computer scientists in different forms: cyberbullying and aggression on Twitter and other media~\cite{,meanBirds,meanBirdsTweb2019,dinakar2011modeling,founta2017large,riadi2017detection}, hateful speech~\cite{waseem2016hateful,davidson2017automated}, and offensive or abusive language~\cite{Chen2012DetectingOffensiveLanguage,lee2002holistic,nobata2016abusive}.
Recently, there were also efforts in the detection of such language in the streaming setting, where online models are trained to detect such behavior~\cite{herodotou2021icde, herodotou2020big-data}.
In the following, we first discuss insights extracted from literature that attempt to model aggression in different ways.
Then, we formally propose the \textit{Aggression Propagation} problem and present how the literature insights can be used to inform existing models into modeling aggression propagation.

\subsection{Aggression Modeling: Literature Insights}
\
\descr {1. \textbf{Influence from strong social relationships.}}\label{sec:insightSocialRelationships}
Aggressive behavior is reactionary and impulsive, and often results in breaking household rules or the law, and can even be violent and unpredictable~\cite{ABDefinition}.
Interestingly, aggressive acts, while reflecting the influence of various mental and physical disorders, in most instances represent learned behaviors from other individuals~\cite{AggressiveBehaviour_Definition}.
In fact, some earlier works proposed that online abusive behavior could be explained using sociology- and psychology-based theories such as \textit{social learning}, \textit{social bonds} and \textit{planned behavior}~\cite{lee2002holistic}.
Furthermore,~\cite{Cheng_2017} observed that a person's negative mood increases the likelihood of adopting negative behavior, which is easily transmitted from person to person.
In addition, \cite{squicciarini2015bully-pairwise} identified that bullies may be able to exert influence through their pairwise interactions with others in the network, and thus spread bullying.
These works lead us to the first insight on aggressive behavior: \textit{due to strong social bonds, users can be influenced by, and learn from others such aggressive behavior.}

\descr {2. \textbf{Influence from social groups.}}\label{sec:insightSocialGroups}
Aggressive adolescents may be unpopular in the larger social community of peers and adults, yet they can be accepted by and closely linked to particular subgroups of peers~\cite{Cairns1988}.
Furthermore, as it was investigated in~\cite{GAM_games}, the personal responsibility exhibited by individuals or groups can be captured by the General Aggression Model (GAM)~\cite{GAM}.
The authors established that when individuals or a group of individuals come to believe either that they are not responsible or that they will not be held accountable by others, the stage is set for the occurrence of violent evil and aggressiveness~\cite{GAM_games}.
In addition, in recent studies on Twitter by Chatzakou et al.~\cite{meanBirds,meanBirdsTweb2019}, cyberbully and aggressive users were found to be less embedded in the network, with fewer friends and smaller clustering coefficient.
Further,~\cite{Kramer8788} found that exposing a person to negative or positive behaviors of those around him, leads him to experience the same emotions as them.
These results lead us to the second insight: \textit{aggressive users may be embedded in small social groups, which can have high impact on their aggression.}

\descr {3. \textbf{Influence due to power difference.}}\label{sec:insightPowerDifference}
Studies have also looked at the emotional and behavioral state of victims of bullying and aggression and how it connects to the aggressor's or victim's network status.
~\cite{corcoran2015cyberbullying} observed that a high power difference in the network status of two individuals can be a significant property of bully-victim relationship.
Also,~\cite{Pieschl2013241} noted that the emotional state of a victim depends on the power of the victim's bullies.
For example, more negative emotional experiences were observed when more popular cyberbullies conducted the attack.
Also, users with higher degree centrality are not only more popular, but potentially more influential on others, since their position in the network offers them greater exposure and thus greater opportunity to have an impact on the behavior of others ~\cite{squicciarini2015bully-pairwise}.
These observations lead us to the third insight: \textit{the power difference that a user may have over another (e.g., due to popularity) can be a decisive factor on the exerted aggression to another user.}

\descr {4. \textbf{Influence due to internal state and external input.}}\label{sec:insightInternalStateExternalInput}
GAM is an integrative approach to understanding aggression that incorporates the best aspects of many domain-specific theories of aggression and takes into account a wide range of factors that affect aggression.
It is separated into two layers, representing the distal causes and proximate causes. 
The distal processes express how biological (e.g., hormone imbalances, low serotonin, and testosterone) and persistent environmental (e.g., difficult life conditions, victimization, and diffusion of responsibility) factors work together to influence a user's personality and increase the likelihood of developing an aggressive personality.
The proximate processes have three stages: (i) inputs, (ii) routes, and (iii) outcomes, that can affect the person's level of aggression and possible reactions to the input.
The reaction that is selected then influences the encounter, which in turn influences the person and situation factors, beginning a new cycle of influence.
The findings from this important study (i.e., GAM) lead us to a forth insight: \textit{users can be influenced by external inputs, but they also try to consolidate them with their internal state of arousal, cognition and affect, before moving to a new state.}

\descr {5. \textbf{Influence can appear in cycles.}}\label{sec:insightInfluenceInCycles}
The overall process outlined by the GAM, and the previously extracted insights (1-4) can be captured in an aggression propagation model.
The user (who could be an aggressor or a victim/normal user) is allowed under monitoring to:
(1) have an internal aggression state of his own,
(2) interact with his neighbors and close friends and receive and/or exert influence of aggression,
(3) assess if he will be changing his stance on aggression, i.e. to become more (or less) aggressive after his interactions,
(4) act by changing (or not) his stance,
(5) repeat these steps in the next cycle (or time step).
This insight allows to build on adapted opinion models that work in simulated rounds or cycles to solve the Aggression Propagation problem, as presented next.

\subsection{Aggression Propagation Problem}

Online users may be ``friends'' and connected in an online social network.
In our setup, user $i$, at time $t$ has its own aggression score that represents his internal, continuous state, $S_{i}^{t}$.
While he interacts with his followers or followings, he may be influenced to be more or less aggressive, thus changing his internal aggression state at every time instance.
The impact that others (i.e., his direct friends or neighborhood) have on his aggression state, can be a function of the strong social relationships with him ($w_{ij}$), his power score $P_{i}$ (e.g., degree centrality), the size of the user's neighborhood ($N_{i}$), etc.
The change of aggression state is continuous, i.e. at every time instance, users are influencing each other's aggression state, partially or totally.
Therefore, the problem of aggression propagation is to model how aggression among users will diffuse or propagate in a network for some time window $W$.
Obviously, this problem has clear similarities with the opinion propagation problem, and techniques to model opinion dynamics could be adapted to model how users influence each other to change aggression state.

As this is the first investigation on this problem, we opt to establish a solid baseline of solutions to the problem at hand, and propose simple, parameter-free models  that are generalizable and applicable to different social networks.
Since aggression can manifest in different ways and intensities, in this work, we study its propagation in a minimal form of one-time aggressive behavior exhibited from one user to another.
Repetition, or accumulation of aggressive behavior (via multiple same-pair interactions) can lead to bullying~\cite{meanBirds,meanBirdsTweb2019}.
Interestingly, this extreme behavior can also be captured in our models with repetitive increase of aggression state.

The following lists the additional notations used in text:

\noindent
\makebox[1.5cm]{$S_{i}^{t}$}			Aggression score of user $i$, at time $t$, $S_{i}^{t}$$\in$$[0, 1]$	\\
\makebox[1.5cm]{$w_{ij}$}			Weight of edge $\big(i, j\big)$ of users $i$ and $j$, defined as		\\
\makebox[1.5cm]{}					Jaccard overlap of neighbor sets: $\frac{ N_{i} \cap N_{j}}{N_{i} \cup N_{j}}$$\in$$[0, 1]$	\\
\makebox[1.5cm]{$P_{i}$}				Power score of user $i$: ratio of in-degree	 over 				\\
\makebox[1.5cm]{}					out-degree: $\frac{inDegree_{i}}{outDegree_{i}} \in [0, 1]$	\\
\makebox[1.5cm]{$A_{x}$}			Selector for applying a factor out of the options:					\\
\makebox[1.5cm]{}					1, $w_{ij}$, $P_{i}$, $P_{j}$, $w_{ij}P_{i}$ and $w_{ij}P_{j}$, for user $x$

Next, we take the five insights identified and embed them in popular opinion models, to construct our proposals for modeling aggression propagation through the network.
Overall, we investigate $26$ propagation models: Voter~\cite{10.1093/biomet/60.3.581}, Deffuant~\cite{doi:10.1142/S0219525900000078}, HK~\cite{hk}, DeGroot~\cite{degroot}, FJ~\cite{fj}, Averaging DeGroot and FJ models, and variants of all these.

\descr {Voter \& Deffuant models \& variants.}\label{voter}
Firstly, we propose four pairwise models based on the Voter model, i.e. one of the simplest and well-known pairwise models in opinion dynamics. 
We assume that after an interaction between two users $i$ and $j$, the aggression score of $i$ changes, because he was influenced (positively or negatively).
The formulation of the first set of proposed models is the following:
\begin{equation}
S_{i}^{t + 1} = A_{j} S_{j}^{t}
\end{equation}
The model names depend on factor $A_{x}$: Voter, \textrm{Voter\_W}, \textrm{Voter\_P} and \textrm{Voter\_WP}.

These models take into account the strong relationship (1st insight) between user $i$ and $j$.
User $i$'s aggression score does not consider its own state but only the aggression score of the neighbor $j$.
The four versions reflect different variations of the Voter model, where the user $i$ assumes the aggression of his neighbor:
1) all of it (i.e., the neighbor is completely affecting the user),
2) weighted by their edge weight (i.e., the neighbor has an influence but only depending on the strength of their relationship),
3) weighted by the Power score of the neighbor (i.e., to capture the concept of power difference that aggressors take advantage of), and
4) weighted by the combination of Power and edge weight.

Another pairwise model suited for interaction within large populations is the Deffuant model that captures confirmation bias, i.e. people's tendency to accept opinions that agree with their own.
To this end, based on Voter and Deffuant models, we propose a second set of models, where user $i$ at $t+1$ does not only take into account the aggression state of his neighbor $j$ (1st insight), but also includes his personal state before making changes in his aggression (4th insight).
Consequently, this set of models can be formalized as follows:
\begin{equation}
S_{i}^{t + 1} = A_{i} S_{i}^{t} + A_{j} S_{j}^{t}
\end{equation}
The model names depend on factor $A_{x}$: \textrm{Deffuant\_W}, \textrm{Deffuant\_P} and \textrm{Deffuant\_WP}.
If factor $A_{x}$ is equal to 1, it is not taken into account.
To maintain the limits of aggression score in a closed interval $[0, 1]$, we normalize the final aggression score of user $i$ using the maximum aggression score from all the neighbors of $i$ at time $t+1$.

\descr{Deffuant \& HK models \& variants.}
Another set of pairwise models we propose rely on the combination of the Deffuant and HK models (in HK, opinions take values in continuous interval, and a bounded confidence limits the interaction of user $i$ with his neighbors), as follows:
\begin{equation}
\textrm{If  } |A_{i} S_{i}^{t} - A_{j} S_{j}^{t}| < d, \textrm{then,   } S_{i}^{t + 1} = A_{i} S_{i}^{t} + A_{j} S_{j}^{t}
\end{equation}
\noindent
The model names depend on factor $A_{x}$: \textrm{HK\_d\_W}, \textrm{HK\_d\_P} and \textrm{HK\_d\_WP}.
It does not include the case of factor $A_{x}$ equal to 1.
This set of pairwise models uses the condition about the bounded confidence limits from the HK model (3rd insight), and updates the aggression score accordingly.
The proposed model is affected by the strong relationship with its neighbor (1st insight) and internal personal state (4th insight) at the previous moment.
We normalize the final aggression score using the maximum aggression score from all of $i$'s neighbors, for those that the treaty is valid at $t+1$.

\begin{table*}[!t]
\caption{Overview of datasets used with our framework.}
\centering
\scalebox{1.20}{
\begin{tabular}{ccc | cc}
				& \multicolumn{2}{c|}{\textbf{Ground truth labeled data (LT)}}	& \multicolumn{2}{c}{\textbf{Unlabeled data (UT)}}	\\
				 \cline{2-5}
				& \textbf{Crawled Net}	& \textbf{Ego-Ego Net (SCC)}			& \textbf{Full Net}	& \textbf{SCC}		\\ \hline
\textbf{\# nodes}	& 1,171,218 			&	314 (57)						& 81,306			& 68,413			\\ \hline
\textbf{\# edges}	& 1,787,543			&	540 (395)						& 1,768,149		& 1,685,163		\\ \hline
\multirow{3}{*}{\textbf{labels}}	&
\multirow{3}{*}{\begin{tabular}[c]{@{}c@{}}Aggressive\\ Normal\end{tabular}}	&
\multirow{3}{*}{\begin{tabular}[c]{@{}c@{}}Bullying\\ Aggressive\\ Normal\end{tabular} $\rightarrow$
\begin{tabular}[c]{@{}c@{}}Aggressive\\ Normal\end{tabular}}	&
\multicolumn{2}{c}{\multirow{3}{*}{\begin{tabular}[c]{@{}c@{}}Aggressive\\ Normal\end{tabular}}} \\
				&				&								&				&				\\
				&				&								&				&				\\ \hline
\end{tabular}
}
\label{table:datasets}
\end{table*}

\descr{DeGroot model \& variants.}
The next set of proposed models take into account the neighborhood of user $i$ for deciding what aggression score to give to the user (2nd insight).
The aggression of a user can be influenced by all of user's neighbors and its internal behavior (4th insight).
As a result, this set of models are variants of DeGroot model (allows a user to consider all or some of their neighbors' score) which considers an average effect across all the neighborhood of the user, and are calculated as follows:
\begin{equation}
S_{i}^{t + 1} = \frac {A_{i} S_{i}^{t} + \sum_{j \in N_{i}}A_{j} S_{j}^{t}}{A_{i} + \sum_{j \in N_{i}} A_{j} }
\end{equation}
\noindent
The model names depend on factor $A_{x}$: Degroot, \textrm{DeGroot\_W}, \textrm{DeGroot\_P} and \textrm{DeGroot\_WP}. 
$A_{x}=1$ corresponds to the original DeGroot model.

\descr{FJ model \& variants.}\label{fjmodels}
We also propose variants of the FJ model (i.e., a variation of the DeGroot model, with the main difference lying in the fact that  in the FJ model each user has an intrinsic initial opinion that remains the same, and an expressed opinion that changes over time), integrating the initial aggression state of an individual in the network (4th insight), along with the user's neighborhood (2nd insight):
\begin{equation}
S_{i}^{t + 1} = \frac {A_{i}S_{i}^{0} + A_{i}S_{i}^{t} + \sum_{j \in N_{i}}A_{j} S_{j}^{t}}{2A_{i} + \sum_{j \in N_{i}} A_{j} }
\end{equation}
\noindent
The model names depend on factor $A_{x}$: \textrm{FJ\_W}, \textrm{FJ\_P}, \textrm{FJ\_WP}. 

\descr{Averaging DeGroot \& FJ models \& variants.}
We propose the following set of models based on DeGroot and FJ, where the aggression score of each user has been inspired by the 2nd, 3rd and 4th insights.
The models are modified by taking the average power score and aggression score from all of user's neighbors, individually:
\begin{equation}
S_{i}^{t + 1} = \big( \frac{A_{i} + \sum_{j \in N_{i}}A_{j}}{1 + \sum_{j \in N_{i}}1} \big) \big( \frac{S_{i}^{t} + \sum_{j \in N_{i}}S_{j}^{t}}{1 + \sum_{j \in N_{i}}1} \big)
\end{equation}
\noindent
The model names depend on factor $A_{x}$: \textrm{avg DeGroot\_W}, \textrm{avg DeGroot\_P}, and \textrm{avg DeGroot\_WP}.

The final set of proposed models is similar to FJ models, but in these we consider the initial aggressive state of each user $i$ (2nd-4th insights).
Thus, the models are as follows:
\begin{equation}
S_{i}^{t + 1} = \big( \frac{A_{i} + A_{i} + \sum_{j \in N_{i}}A_{j}}{1 + 1 + \sum_{j \in N_{i}}1} \big) \big( \frac{S_{i}^{0} + S_{i}^{t} + \sum_{j \in N_{i}}S_{j}^{t}}{1 + 1 + \sum_{j \in N_{i}}1} \big)
\end{equation}
\noindent
The model names depend on factor $A_{x}$: \textrm{avg FJ\_W}, \textrm{avg FJ\_P}, and \textrm{avg FJ\_WP}. 

Next, we explain how all presented models are implemented into a simulator for exhaustive experimentation with different parameter settings using real Twitter data.

\section{Simulation Methodology}\label{sec:overview}

\begin{figure*}[htbp]
\begin{center}
	\includegraphics[scale=0.6]{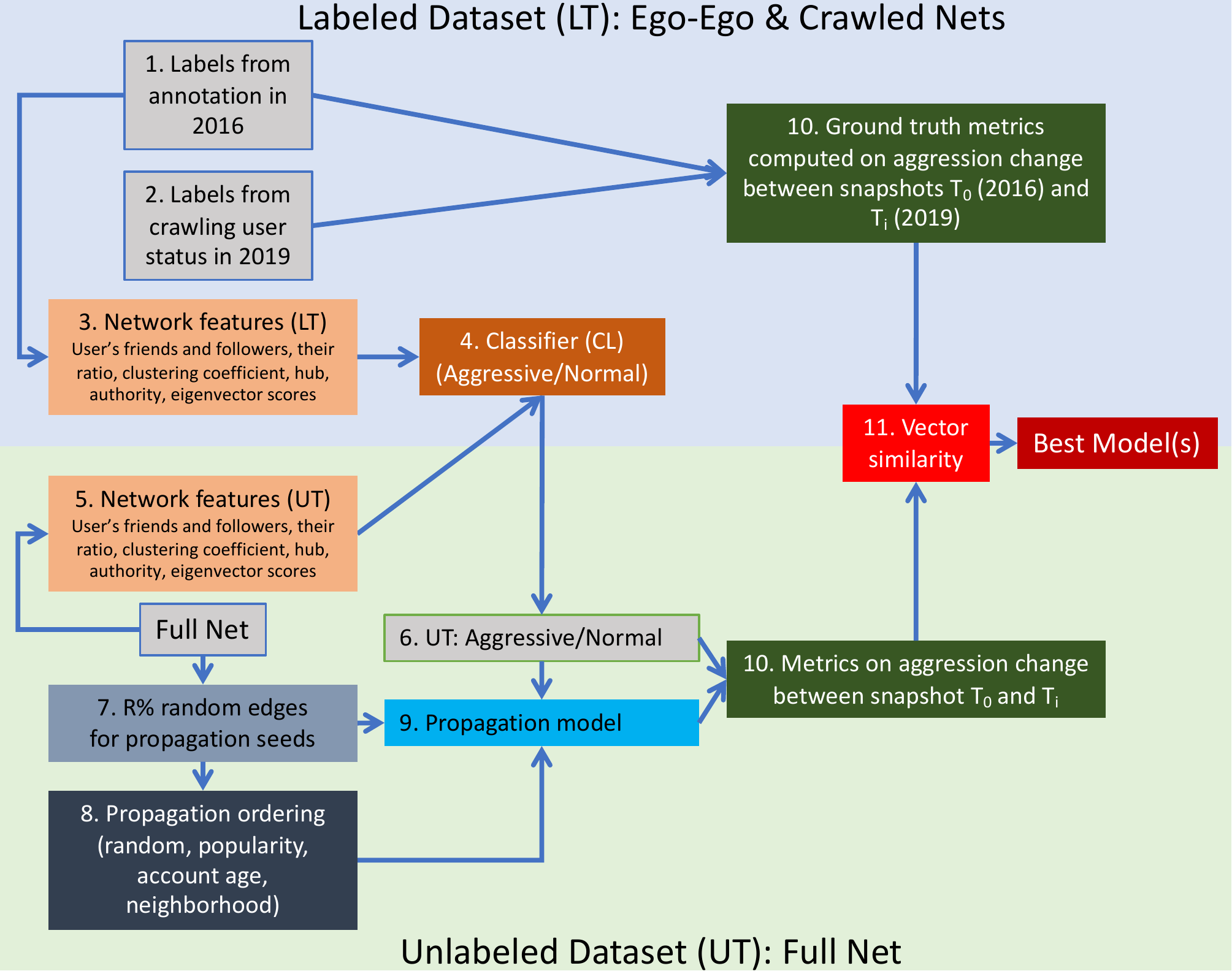}
	\caption{Overview of the simulation methodology.}
	\label{fig:methodology}
\end{center}
\end{figure*}

In this section, we outline the methodology followed to simulate propagation of aggression in a social network, given each one of the models proposed earlier (See Figure~\ref{fig:methodology} for an overview).
First, some users in the network are assumed to be aggressive, and the rest as normal, formalizing the network's initial state.
As time passes, users interact with, and may affect, each other to become more or less aggressive, thus changing the overall state of aggression of the network through time.
Different models can be used to describe these user interactions and aggression change.
To identify which model is better for this task, we compare each model's imposed aggression changes with real (ground truth) data of aggression propagation.
Each model performs differently through the simulation, and may match best with the ground truth data at different point in the simulated time.
Therefore, at regular time intervals during each simulation we capture snapshots of the network's aggression state and compare each model with the validation data.

Next, we address in the simulator design important factors that can impact the exploration of this complex problem:\\
\indent 1. Online social network \& ground truth data\\
\indent 2. Aggressive and normal users\\ 
\indent 3. Users (edges) to perform propagation\\
\indent 4. Ordering of users to perform propagation\\ 
\indent 5. Propagation models applied to modify users' scores\\
\indent 6. Metrics used to capture (change of) state of aggression\\
\indent 7. Metrics to compare state of aggression in simulated and\\\indent validation networks

\subsection{Online social network \& ground truth data}

For our experimental exploration, two main datasets are considered, as presented in Table~\ref{table:datasets}.
Firstly, a Labeled Twitter ($LT$) network was constructed using $401$ annotated users from~\cite{meanBirds} as seeds for crawling their ego-network, to record their friends and followers.
This set of data constitutes our ground truth, as explained next.
We performed two crawls of these $401$ users' ego networks: in 06-09/2016 and in 04-05/2019.
These $401$ users were originally annotated in 2016 as \textit{normal}, \textit{aggressive}, \textit{bullying}, or \textit{spammers} (Step 1, Fig.~\ref{fig:methodology}).
We removed the spam class and merged all aggressive and bullying users under the \textit{aggressive} label.
Thus, the $401$ users have two labels for 2016.
Their friends and followers do not have an aggression label.
However, since during the crawling they were found as active, we make the conservative assumption that they were normal (even though this is a lower bound on aggressive labeling).

When the re-crawl of these users in 2019 was completed (Step 2, Fig.~\ref{fig:methodology}), $\sim$$650$K were found to be active, $\sim$$30$K were found suspended, $\sim$$37$K had deleted their account.
For $\sim$$452$K users the status was not retrievable.
We make the assumption that active and deleted users in 2019 can be considered \textit{normal}, and suspended users are \textit{aggressive}, since at some point in the past they violated Twitter rules.
We justify this assumption with the following reasoning.
The authors in~\cite{meanBirdsTweb2019} classified users as \textit{aggressive} or \textit{normal} based on their activity, account and network characteristics, with high confidence based on crowdsourced annotations.
When they examined the account status of these $401$ users after regular time periods (in 2017 and 2018), many of the \textit{aggressive} users were suspended by Twitter, thus justifying their original annotation.\footnote{Twitter suspends an account (temporarily or even permanently, in some cases) if it has been hijacked/compromised, is considered spam/fake, or if it is abusive.}
Therefore, the characterization of the re-crawled users as \textit{aggressive} or \textit{normal} based on their account status can be considered realistic.
To this end, the overall \textbf{Crawled Network $LT$} has $1,171,218$ users (including the $401$ seeds) and $1,787,543$ edges connecting them.
Indeed, we note this dataset reflects only the ego-networks of $401$ users, i.e. it is not a complete social network of 1.1M users.
Also, $477,113$ edges were dropped, since their endpoints' status was not retrievable.
In both crawls of 2016 and 2019 for this network, these users are labeled as \textit{normal} or \textit{aggressive}, following the above process.

\begin{table*}[!t]
\centering
\caption{Metrics used to measure aggression change.}
\scalebox{1.0}{
\begin{tabular}{cp{9.7cm}l}
\textbf{Metric of state at time $t$=$i$}					&\textbf{Explanation}									\\ \hline
\{n$\}_i$ $||$ \{a$\}_i$								& portion of normal ($||$ aggressive) users in the network					\\
\{N-N$\}_i$ $||$ \{A-A$\}_i$							& portion of edges that both users $i$ and $j$ are normal ($||$ both users are aggressive)		\\
\{N-A$\}_i$ $||$ \{A-N$\}_i$							& portion of edges that user $i$ is normal \& $j$ aggressive ($||$ user $i$ is aggressive \& $j$ is normal)	\\
\textbf{State change between $t$=$i$ and $t$=$j$}			&\textbf{Explanation}									\\ \hline
\{n$\}_i$ $\rightarrow$ \{n $||$ a$\}_j$					& portion of normal users in the initial state who remain normal or become aggressive, respectively\\
\{a$\}_i$ $\rightarrow$ \{n $||$ a$\}_j$					& portion of aggressive users in the initial state who become normal or remain aggressive, respectively\\ \hline
\{N-N$\}_i$ $\rightarrow$ \{N-N $||$ N-A $||$ A-N $||$ A-A$\}_j$	& portion of edges that users $i$, $j$ were normal, and remain normal, or one / both become aggressive, respectively \\
\{N-A$\}_i$ $\rightarrow$ \{N-N $||$ N-A $||$ A-N $||$ A-A$\}_j$	& same as above for edges where $j$ is aggressive at the initial state	\\
\{A-N$\}_i$ $\rightarrow$ \{N-N $||$ N-A $||$ A-N $||$ A-A$\}_j$	& same as above for edges where $i$ is aggressive at the initial state	\\
\{A-A$\}_i$ $\rightarrow$ \{N-N $||$ N-A $||$ A-N $||$ A-A$\}_j$	& same as above for edges where both $i$ and $j$ are aggressive at the initial state \\ \hline
\end{tabular}}
\label{table:metrics}
\end{table*}

Secondly, an \textbf{Unlabeled Twitter} ($UT$) network was considered, with $81,306$ users, and $1,768,149$ directed edges between them, acquired from~\cite{mcauley12social-circles}.
We specifically focus on the network's Strongly Connected Component (SCC) and the final $UT$ network has $68,413$ users and $1,685,163$ edges, with properties of a proper social network, in contrast to the LT ego-networks.
Also, and following past work on network analysis~\cite{zuo14indirectties}, we apply weights on the edges based on the Jaccard overlap of social circles between two users $i$ and $j$, $w_{ij} = \frac{ N_{i} \cap N_{j}}{N_{i} \cup N_{j}} \in [0, 1]$.
As explained next, we apply a classification process to characterize these unlabeled users of this dataset as either \textit{aggressive} or \textit{normal}, using the $401$ labeled users from $LT$.

\subsection{Which users should be aggressive?}\label{sec:4.2}
In the $UT$, we have no labels of aggression, and thus, we do not know which users exhibit aggression and which are normal.
An ideal process for labeling users as aggressive or not would be to subject this dataset to annotation by use of crowdsourcing or experts, as done in past studies~\cite{meanBirds}.
However, this process is expensive, slow and non-scalable~\cite{founta2017large}.
It also assumes that there is access to each user's activity (i.e., tweets) to judge if they are behaving aggressively or not.
Therefore, and in order to identify users who should be labeled as aggressive, we proceed as follows:
First, we use the ego-network data of the $401$ users from $LT$, to train a classifier (\CL~=~Random Forest\footnote{Number of trees generated: 10; Maximum depth set: unlimited}) on users' network features\footnote{User's friends and followers, their ratio, user's clustering coefficient score, hub score, authority score, and eigenvector score.}, and thus to infer the likelihood that a user will be aggressive (Steps 3 and 4, Fig.~\ref{fig:methodology}).
Overall, by training and testing the \CL on the $LT$'s $401$ annotated users using repeated 10 cross-fold validation (i.e., train on 90\% of users and test on 10\%), we achieved $93.24\%$ Accuracy and $93.2\%$ Precision and Recall.
Second, we extract the same network properties from the $UT$ network (Step 5, Fig.~\ref{fig:methodology}), and apply \CL on them, in order to label $UT$'s users as \textit{aggressive} or not, based on some threshold (Step 6, Fig.~\ref{fig:methodology}).
Thus, we transfer the label ``knowledge'' or ``learning" from the past labeled network ($LT$) to the unlabeled one ($UT$), using only network features which are present in both datasets.

By applying \CL on the $UT$ network, we got $8.5\%$ or $5,820$ users who were labeled as aggressive.
We verified that users selected to be aggressive (or normal) in $UT$ had similar distributions for their network properties with the $401$ annotated users in $LT$.
Each aggressive (normal) user $i$ was given a score $S_i=1$ ($S_i=0$).
Interacting users can modify each other's aggression state, leading to users with scores between 0 and 1.

\subsection{Users to perform propagation}
We select a set $\mathbb{R}$ of random users for executing the propagation model (Step 7, Fig.~\ref{fig:methodology}).
A large $\mathbb{R}$ can cover a larger portion of the network, but can be extremely costly to simulate.
We opt for $10\%$ of random edges to make the problem computationally tractable, covering $65.3\%$ of total users of the $UT$ network.
For completeness, we also tried for limited number of experimental configurations the full network (i.e., $100\%$ of edges) and the results were similar.

\subsection{Propagation changes applied}\label{sec:orderingofchanges}

Users in $\mathbb{R}$ that were selected for propagation may interact with each other in different ways (Step 8, Fig.~\ref{fig:methodology}):
\vspace{-1mm}
\begin{enumerate}
\item Randomly, i.e. the selected users are randomly shuffled before their aggression is propagated.
\item Based on the most popular (or least popular) user (e.g., using their degree centrality).
\item Based on the neighborhood involved (i.e., group users based on neighborhood and propagate between them).
\item Based on their network id (a proxy of their account age, a trend confirmed in our data).
\end{enumerate}
\vspace{-1mm}
We measure how each method impacts the aggression change of each model during simulation.

\subsection{Propagation models used}
A propagation model can be used to simulate how information or influence or type of behavior can propagate from one user to another, and consequently from one part of the network to another.
The model used may have parameters to be tuned, or can also be parameter-less.
In total, we consider all models and variants proposed in Section~\ref{sec:models} (Step 9, Fig.~\ref{fig:methodology}).
All are parameter-less, making them simple and generalizable on different networks and setups.

\subsection{Metrics used to measure aggression change}
We measure the state of aggression of users and network, and how it changes through simulated time using $26$ different metrics, as explained in Table~\ref{table:metrics}.
These elements capture the state of the network with respect to users and edges and their label at time $t=i$, and how they changed through the simulation between time $t_i$ and $t_j$ (declared as $\rightarrow$).
Each of these metrics can be computed at regular snapshots ($T_0$, $T_1$, $T_2$,~$\dots$,~$T_N$), by comparing the network state at a given snapshot ($T_N$) with initial state $T_0$.

\subsection{Measuring ground truth metrics}\label{sec:groundtruth}

In the previous paragraphs, we described the list of metrics we measure to quantify the changes in state of aggression in the network.
We compute these metrics in the $LT$ network using the previously mentioned time crawls (2016 and 2019).
Thus, we construct a ground truth or validation vector for the above mentioned $26$ metrics (Table~\ref{table:metrics}), which capture the change of aggression of users and type of edges (A-A, A-N, etc.)
(Step 10, Fig.~\ref{fig:methodology}).
However, we do not take into account the following metrics: n, n $\rightarrow$ {n}, N-N, and N-N $\rightarrow$ {N-N}, as they are dominant and overtake the rest of (evidently smaller) changes.

\subsection{Comparing simulation \& ground truth data}\label{sec:comparesimandgroundtruth}
The above set of metrics is computed for all models and for 10 time snapshots per simulation (Step 10, Fig.~\ref{fig:methodology}).
Using a pre-selected threshold $T_A$ for each user's aggression score, we binarize their final state and thus, compute overall aggression change in nodes and edges.
Then, we compare with the validation vector from the ground truth data (Step 11, Fig.~\ref{fig:methodology}).
This comparison is executed using standard similarity metrics such as Cosine similarity, Pearson correlation, Spearman rank correlation, and Euclidean distance.
This comparison establishes how close a model changes the state of aggression of the network (in both nodes and edges) to match the ground truth.

\section{Analysis of Simulation Results}\label{sec:analysis}

In this section, we show the results from the extensive ($>$5k) simulations performed, under different experimental settings used:
26 propagation models,
10 thresholds for $T_A = 0.05,~\dots,~0.9$, 
comparisons with 10 time snapshots, 
4 metrics for comparing ground truth with model performance in each snapshot, 
5 types of orderings of users to propagate aggression, 
and 10\% random edges and their users.
Note: We also tested the full network (i.e., $100\%$ of edges) for a limited number of experimental configurations, and the results were similar to using $10\%$ of random edges.
Therefore, and to reduce computational overhead (each experimental setup on the full network takes 2-3 days to complete) we performed more in-depth experiments using the 10\% of random edges.

\subsection{Which models are stable and perform best?}\label{sec:modelsperformance10}

\begin{figure*}[th]
\begin{center}
	\subfigure[\textit{$T_A=0.3$}]{\includegraphics[width=8cm]{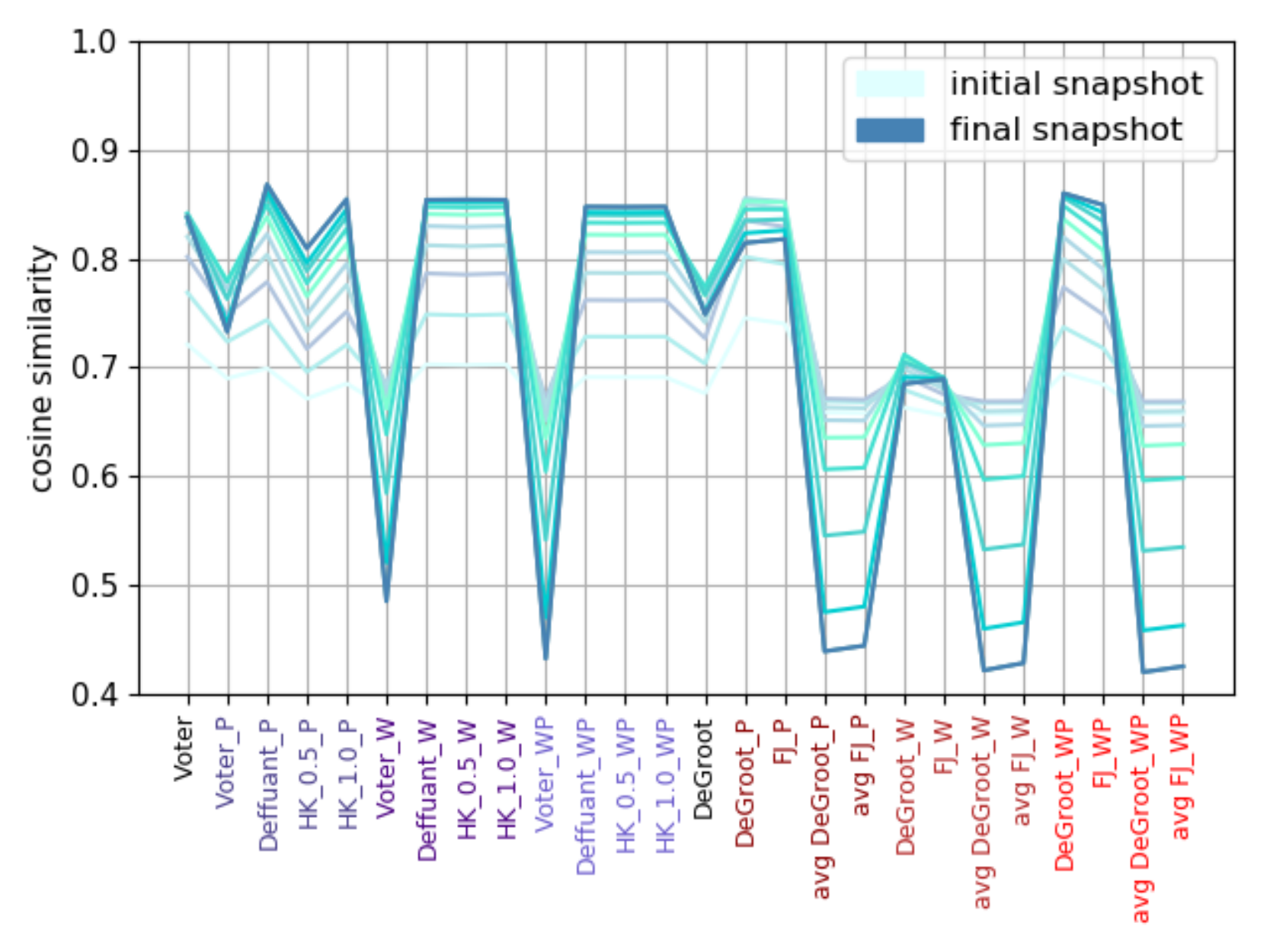}}
	\subfigure[\textit{$T_A=0.5$}]{\includegraphics[width=8cm]{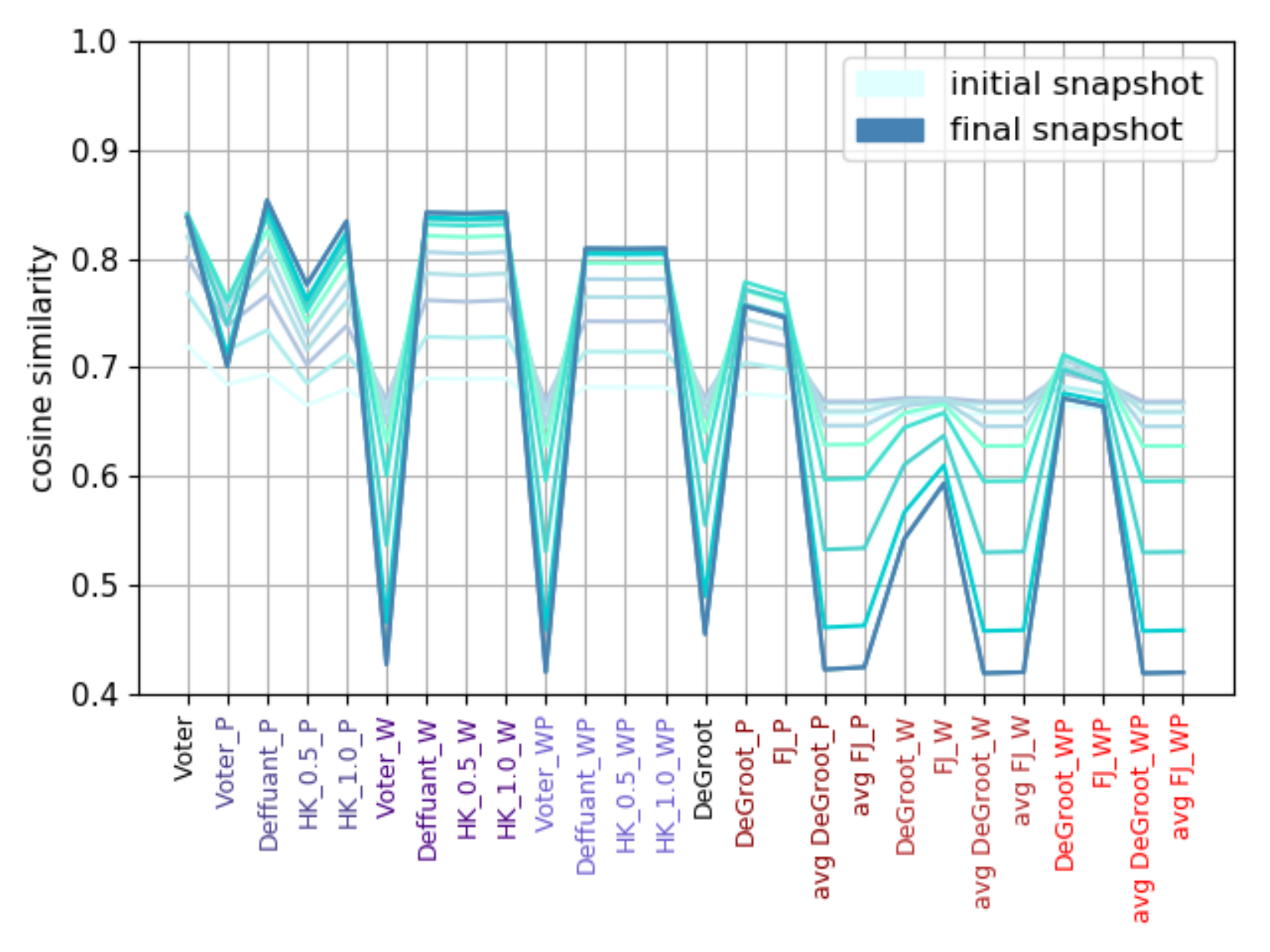}}
	\subfigure[\textit{$T_A=0.7$}]{\includegraphics[width=8cm]{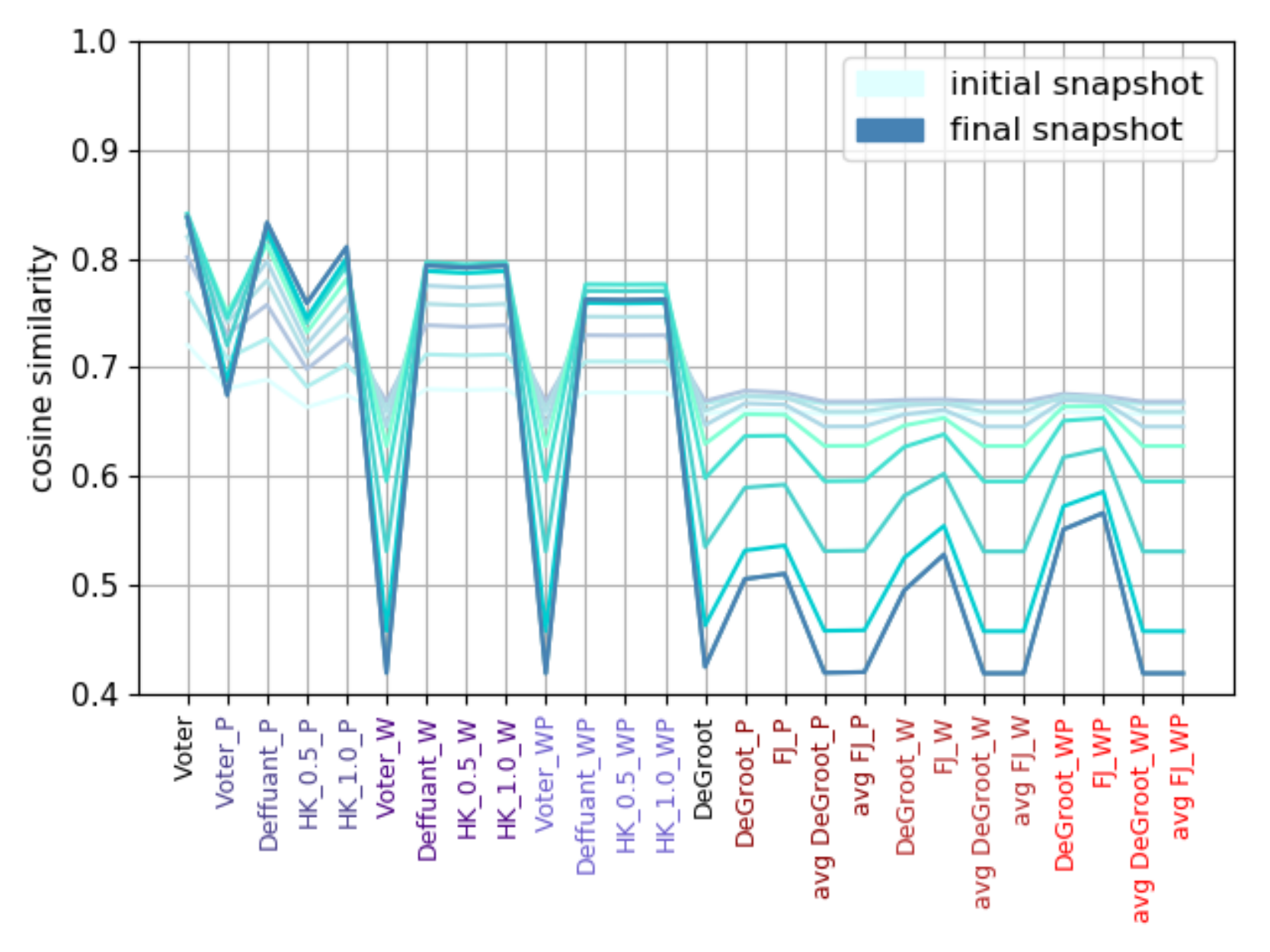}}
	\subfigure[\textit{$T_A=0.9$}]{\includegraphics[width=8cm]{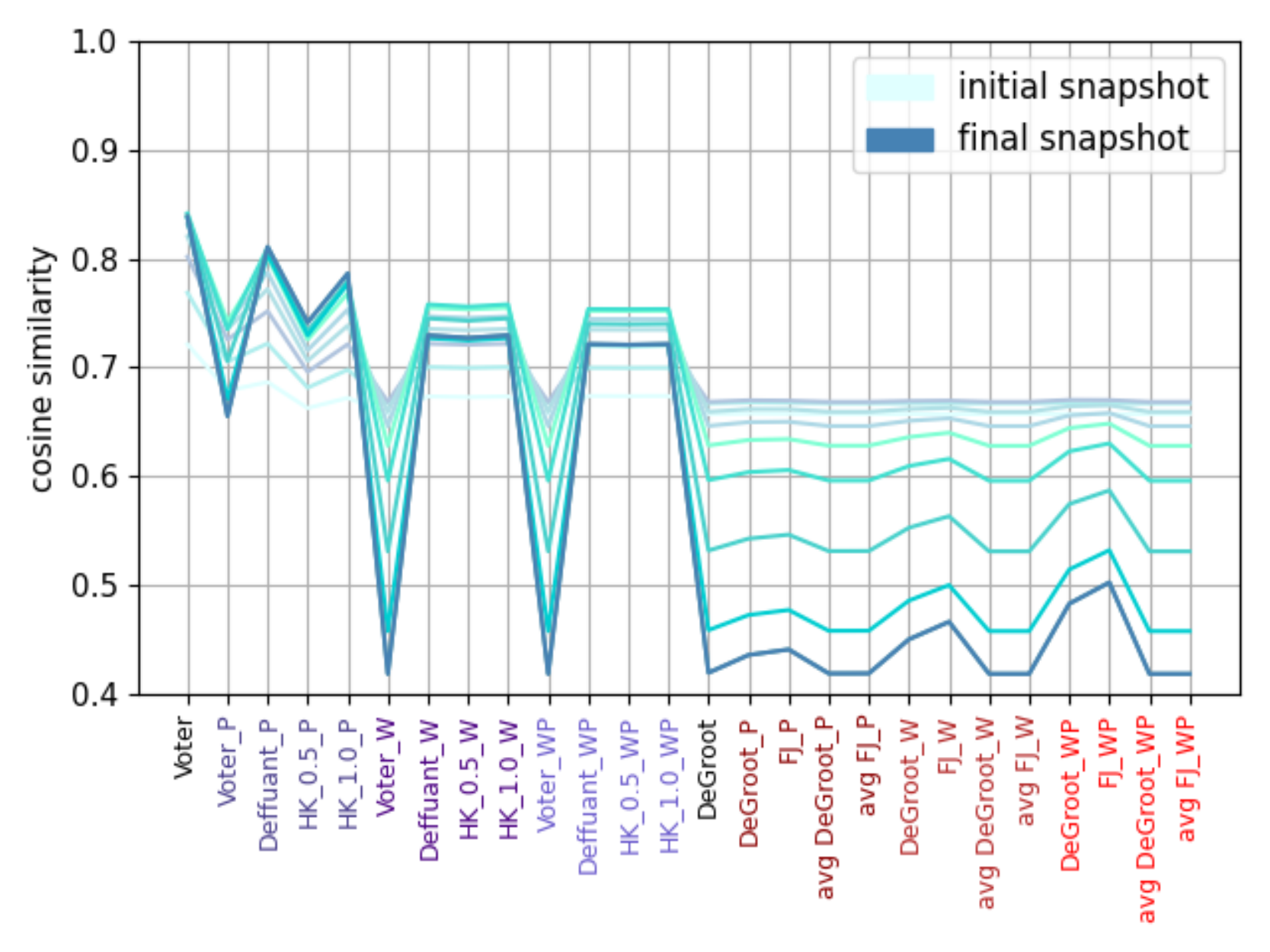}}
	\caption{Cosine similarity of all proposed models with the validation vector, for 10\% of selected edges and different thresholds $T_A$.
	The different time snapshots of the simulation are colored from light to dark blue.
	We group the models in sets based on their commonality.}
	\label{fig:cosineSimilarityPlots}
\end{center}
\end{figure*}

\descr{Different aggression thresholds $\boldsymbol{T_A}$:} 
The first step towards analyzing the simulation results is to compare the proposed models with respect to their performance.
Cosine similarity is used for the comparison, with the threshold above which a user is characterized as aggressive or not, to be varying between $T_A~$=$~0.05,~\dots,~0.9$.
Figure~\ref{fig:cosineSimilarityPlots} plots the cosine similarity for all considered models in relation to the validation vector of real data, for four threshold values as examples ($T_A=0.3$, $0.5$, $0.7$ and $0.9$).
The results for the rest of the thresholds are omitted for brevity.
Overall, we observe models showing stable (low or high) performance, independent of $T_A$ selected, or models highly depended on $T_A$.
Specifically, \textit{Deffuant\_P}, \textit{Voter}, and \textit{HK\_*\_P} achieve high performance (similarity~$>$~0.8), regardless of the $T_A$ selected.
On the contrary, \textit{Voter\_W}, \textit{Voter\_WP}, and \textit{Averaging\_*\_*} show lower performance (similarity~$<$~0.7), irrespective of $T_A$ selected.
Alternatively, there are models whose performance highly depends on the change of $T_A$.
\textit{Deffuant\_W*} and \textit{HK\_*\_W*} fluctuate from average to high performance (0.7~$<$~similarity~$<$~0.85), and \textit{Degroot*} and \textit{FJ\_*} show highly varying performance (0.45~$<$~similarity~$<$~0.85).
Based on the comparison across all thresholds, we conclude that $T_A=0.5$ is a reasonable threshold and we use it for the rest of our experiments.

\descr{Performance within models:}
Fixing $T_A=0.5$, we observe that with \textit{Deffuant\_P}, we achieve best performance.
Also, $Voter$, \textit{Deffuant\_W}, and \textit{HK\_*\_W} are among the top models.
We note that when the edge weight ($W$) is considered, the performance in some cases is adversely affected.
For example, the \textit{Voter\_W} model reaches similarity with ground truth less than $0.5$.
Mixed results are observed when the Power score ($P$) is used.
For example, in \textit{Deffuant} and \textit{DeGroot} models, the performance increases, indicating that the influence of user $j$ to its neighbor $i$ is more important when it is not constrained to the given edge $(i,j)$, but instead, when the neighborhood of $i$, $N_i$, is considered.
\textit{Degroot} and $FJ$ models perform similarly, indicating that the state of neighborhood of the user has no significant influence to the overall performance.
Finally, averaging \textit{DeGroot} and $FJ$ models perform the worst, regardless of whether the edge weight or power score are considered, separately or in combination.

\begin{figure*}[!t]
\begin{center}
	\subfigure[\textit{Voter}]{\includegraphics[width=8cm]{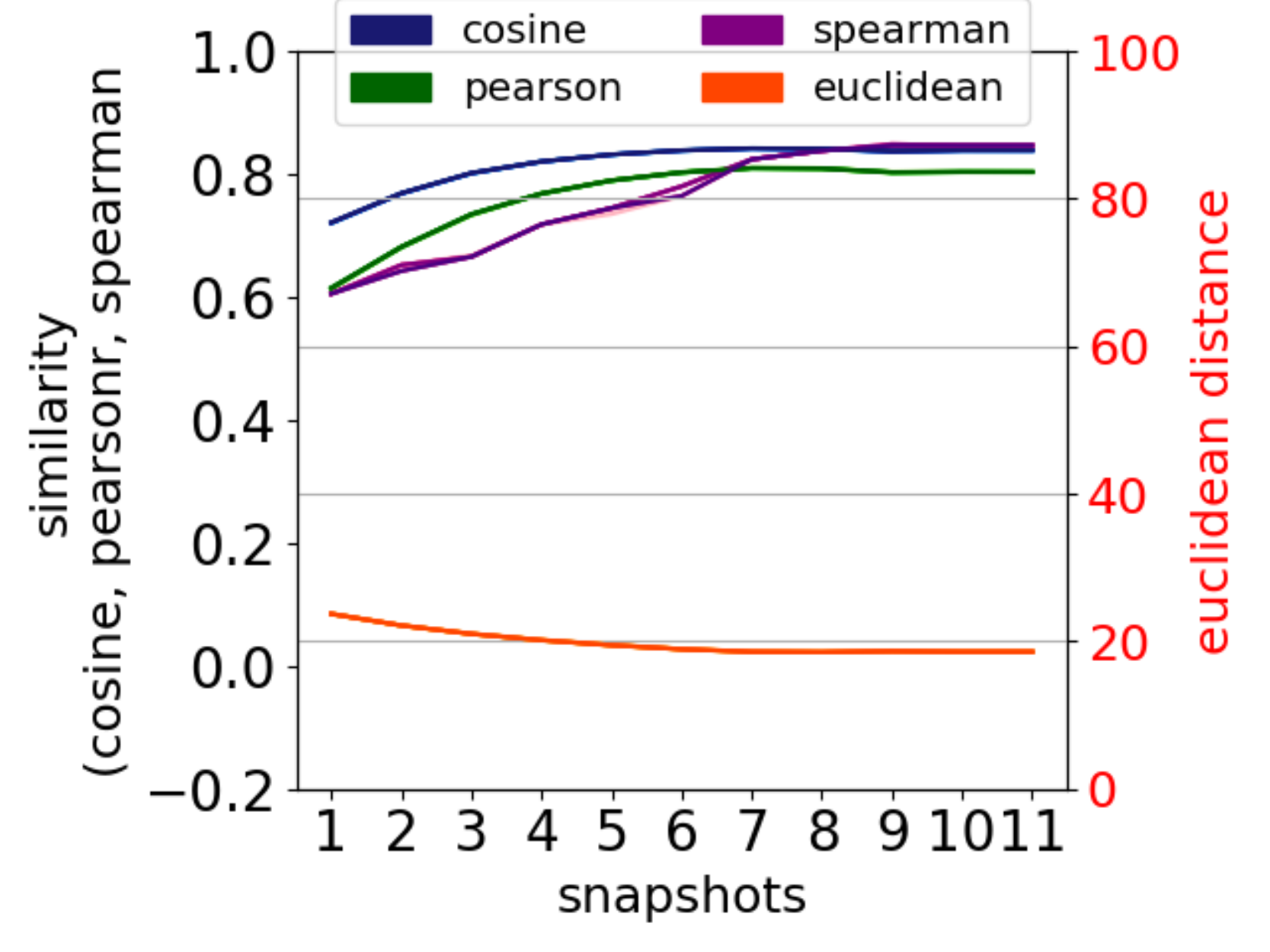}} 
	\subfigure[\textit{Deffuant\_P}]{\includegraphics[width=8cm]{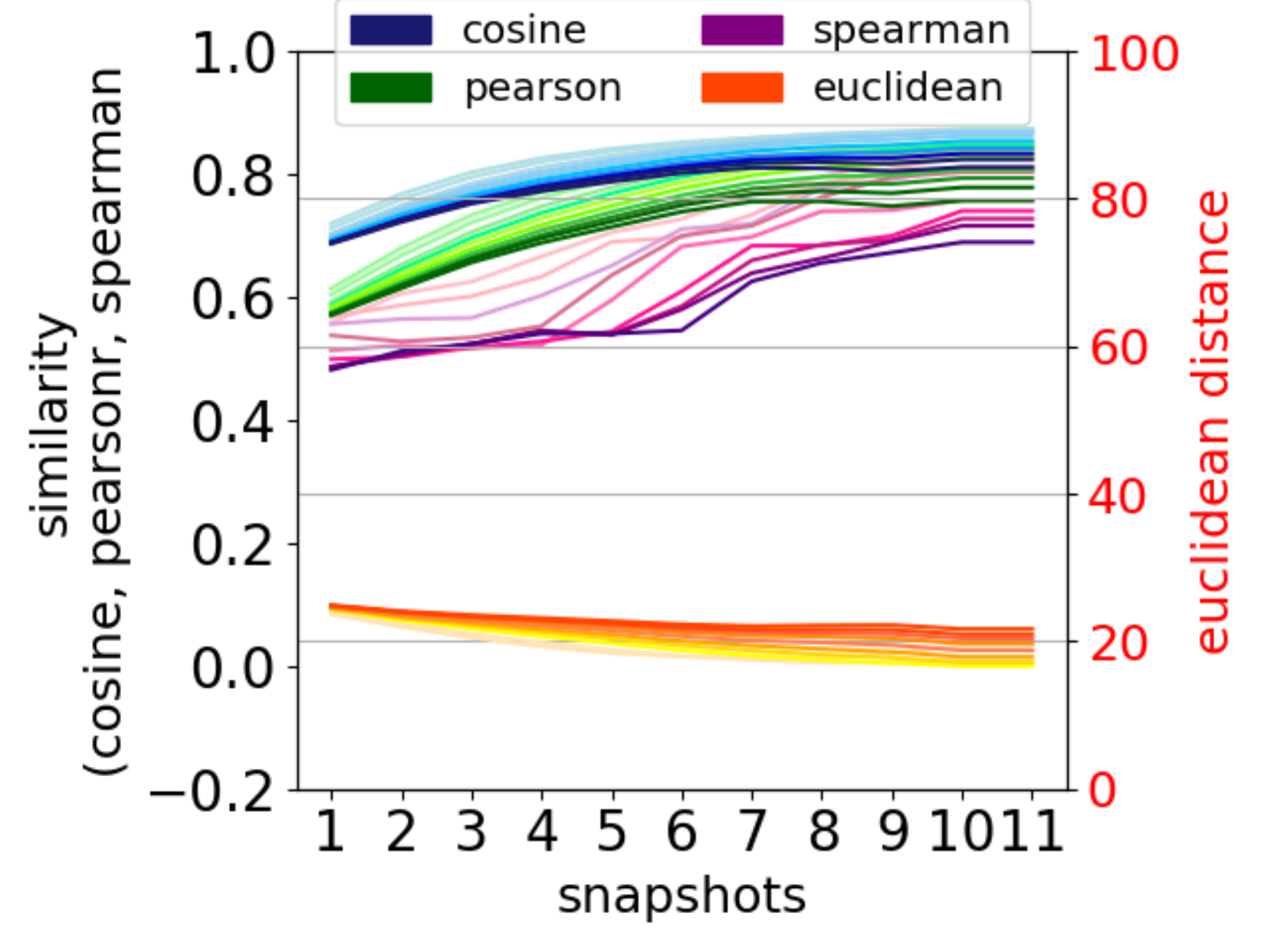}} 
	\subfigure[\textit{HK\_0.5\_P}]{\includegraphics[width=8cm]{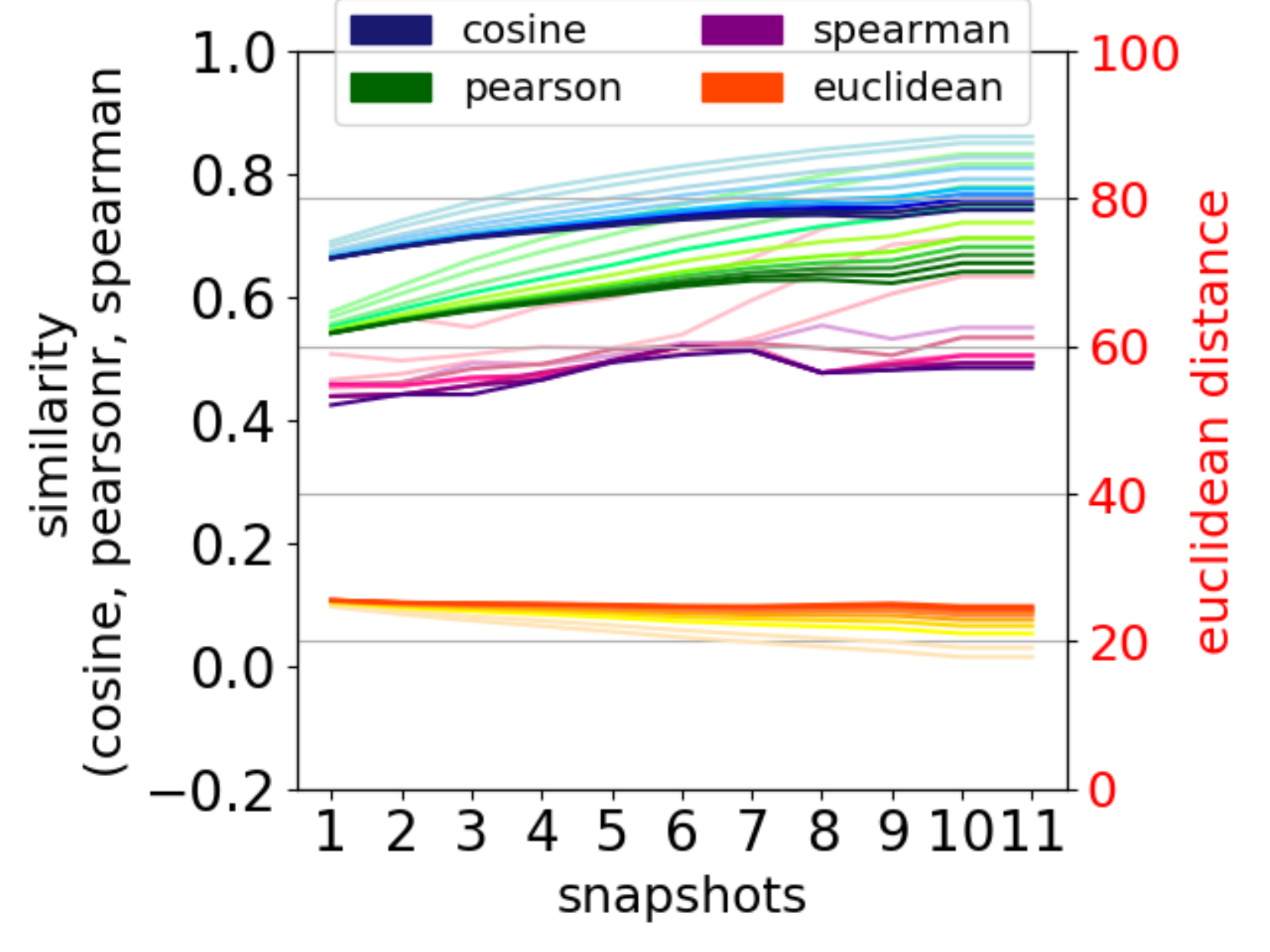}}
	\subfigure[\textit{HK\_1.0\_P}]{\includegraphics[width=8cm]{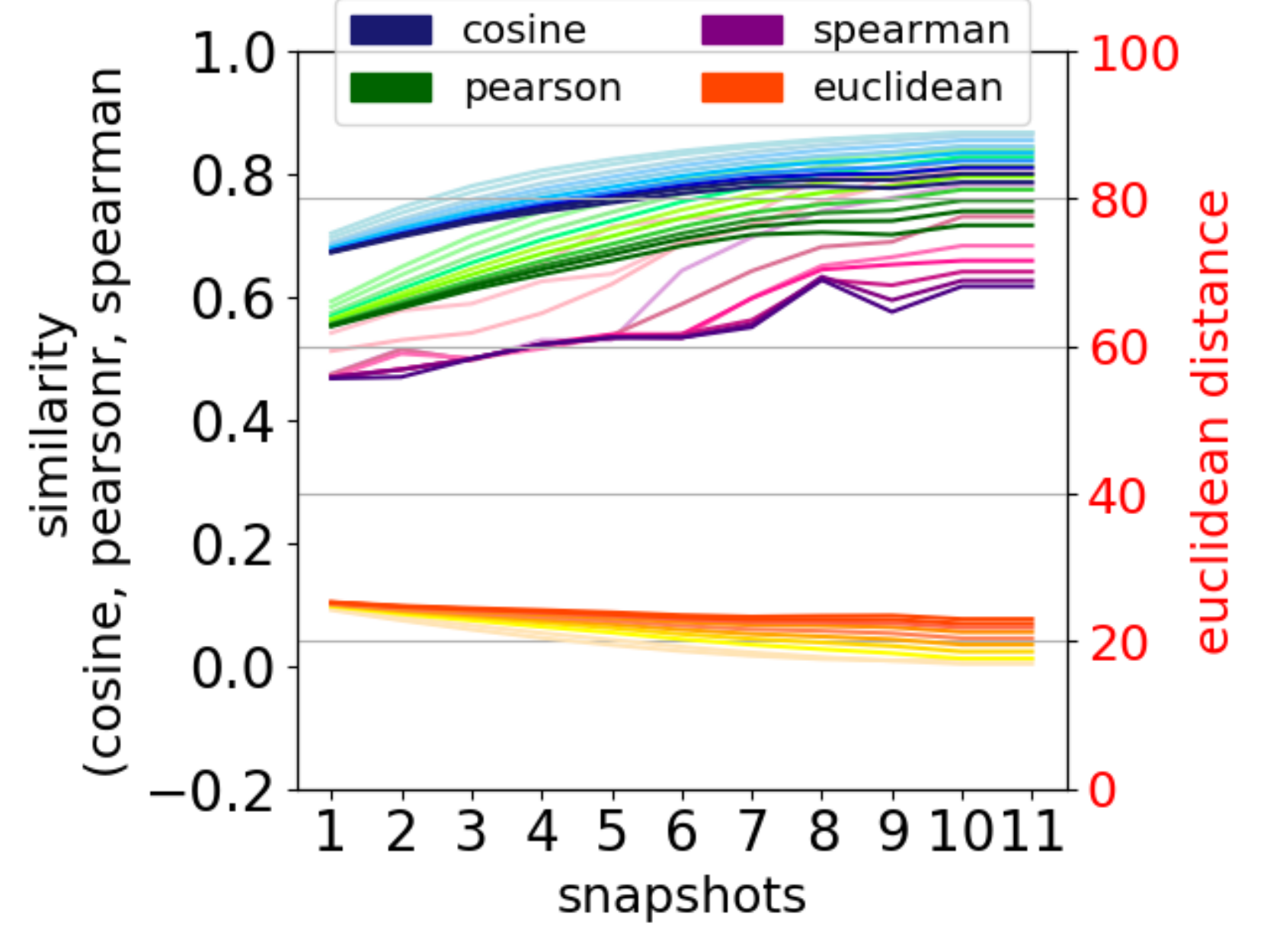}}
	\caption{Similarity of the top 4 performing models with ground truth, through 10 time snapshots: (a) \textit{Voter}, (b) \textit{Deffuant\_P}, (c) \textit{HK\_0.5\_P}, and (d) \textit{HK\_1.0\_P}.
	We show 4 similarity metrics: cosine similarity, Pearson correlation, spearman correlation, and euclidean distance.
	Variation of colors for a metric illustrate the performance of the given model and metric for different thresholds $T_A=0.05, \dots, 0.90$.}
	\label{fig:similarityTypeDPer10BiasedTop4Models}
\end{center}
\end{figure*}

\descr{Takeaways.}
From the top performing models, i.e., \textit{Deffuant\_P}, \textit{Voter}, and \textit{HK\_*\_P}, two main observations can be made: (i) a user's internal aggression state is highly dependent on their mate's aggression state (i.e., with whom there is a direct interaction / relationship), and (ii) the internal aggression state of a user constitutes an important factor in aggression propagation.
This observation aligns well with the 4th insight in Sec.~\ref{sec:insightInternalStateExternalInput}.
In situations with various options for reaction, the inner state (in our case, the aggression state) of individuals, as well as those with whom there is a direct connection with, are key factors in the subsequent state of the individuals themselves~\cite{hatfield1993emotional}.
This is also reflected by the top models, as, on the one hand, they are all pairwise models, while on the other, apart from $Voter$, they consider a user's internal aggression state before making aggression changes.
Overall, based on the best models, we observe that online aggression (especially when taking place on Twitter) is propagating from one user to another; users are not so influenced by their neighborhood.
Aggressive users have been shown to be less popular (i.e., smaller number of followers and friends) than normal users~\cite{meanBirdsTweb2019} which could explain the fact that aggressive users are more affected by direct relationships rather than their neighborhood aggression state.

\subsection{How do models perform over time?}\label{sec:differentsnapshots10}

Next, we examine how model performance is affected when different time snapshots of the network's aggression state are considered.
This analysis is done to: (i) compare with ground truth each model's state at progressing simulation times, and (ii) detect which of the snapshots was better fitting the real data.
We remind the reader that there is no point in simulating propagation until the models converge to some steady state, since the time taken for this may not match the timing the ground truth data were captured.
For our analysis, we focus on the top four performing (regardless of $T_A$) models, i.e., \textit{Deffuant\_P}, \textit{Voter}, \textit{HK\_0.5\_P}, and \textit{HK\_1.0\_P}.

Figure~\ref{fig:similarityTypeDPer10BiasedTop4Models}
shows that for all models, performance is lower within the first snapshots and gradually increases, to stabilize in the last snapshots.
As for the similarity metrics, they follow a similar pattern across models, indicating that either of them can be used to do the performance analysis, and that the comparison results are stable against analysis that considers ranks or absolute numbers in the vectors compared.

\descr{Takeaways.}
These models successfully capture how a network's aggression status changes across time.
The notion of snapshots constitutes a valid process in representing the aggression propagation, since it captures the way aggression is (or will be) expected to be in real time in a network.
By focusing on user interactions (\textit{Voter} model), or user's internal aggression state (\textit{Deffuant\_P}, \textit{HK\_0.5\_P}, and \textit{HK\_1.0\_P}), the aforementioned models can be used to track how aggression propagates in networks with similar properties.

\subsection{Is the order of changes important?}\label{sec:order-of-change}

\begin{figure*}[!t]
\center
    \subfigure[Voter]{\includegraphics[width=8cm]{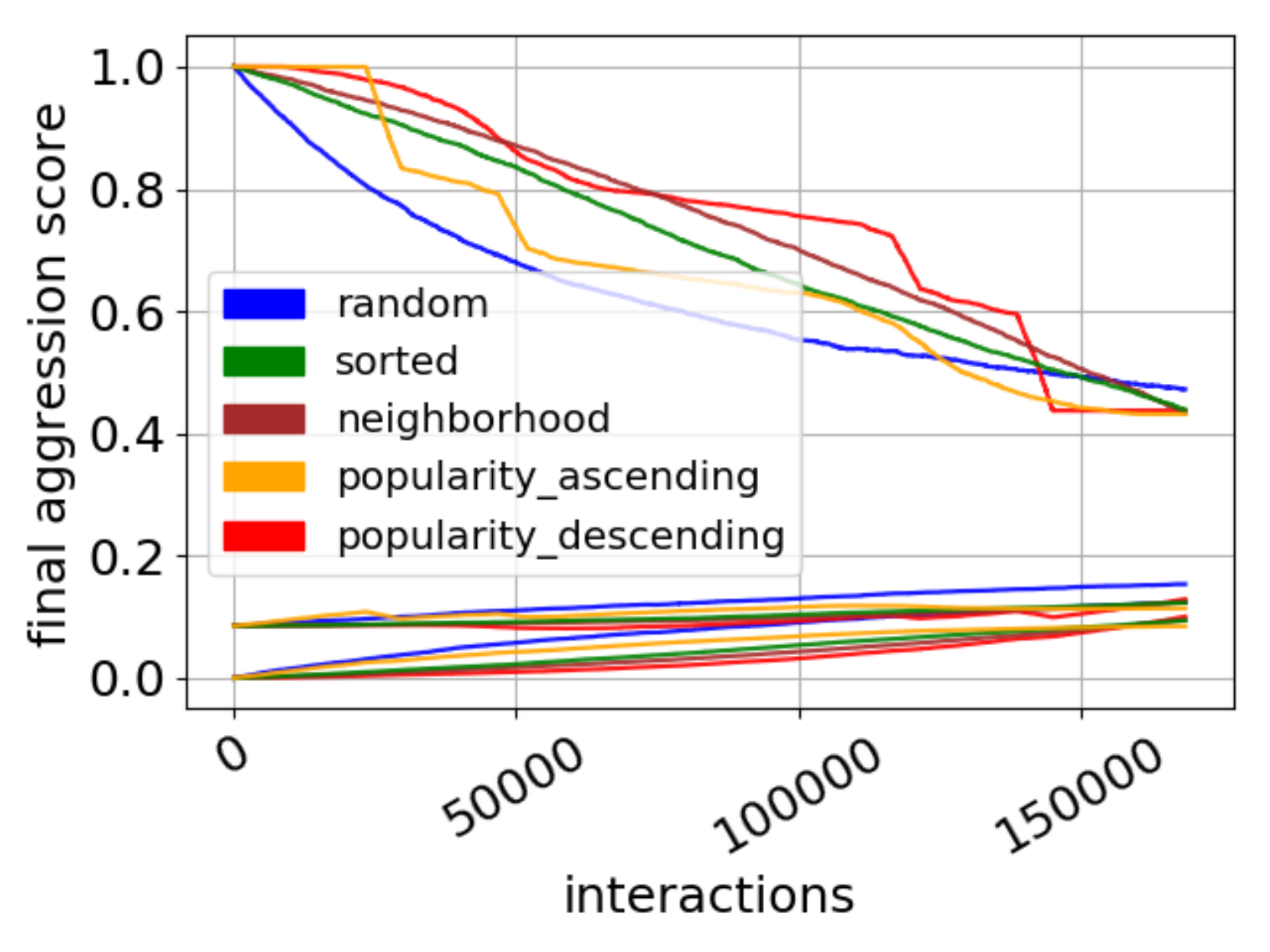}} 
    \subfigure[Deffuant\_P]{\includegraphics[width=8cm]{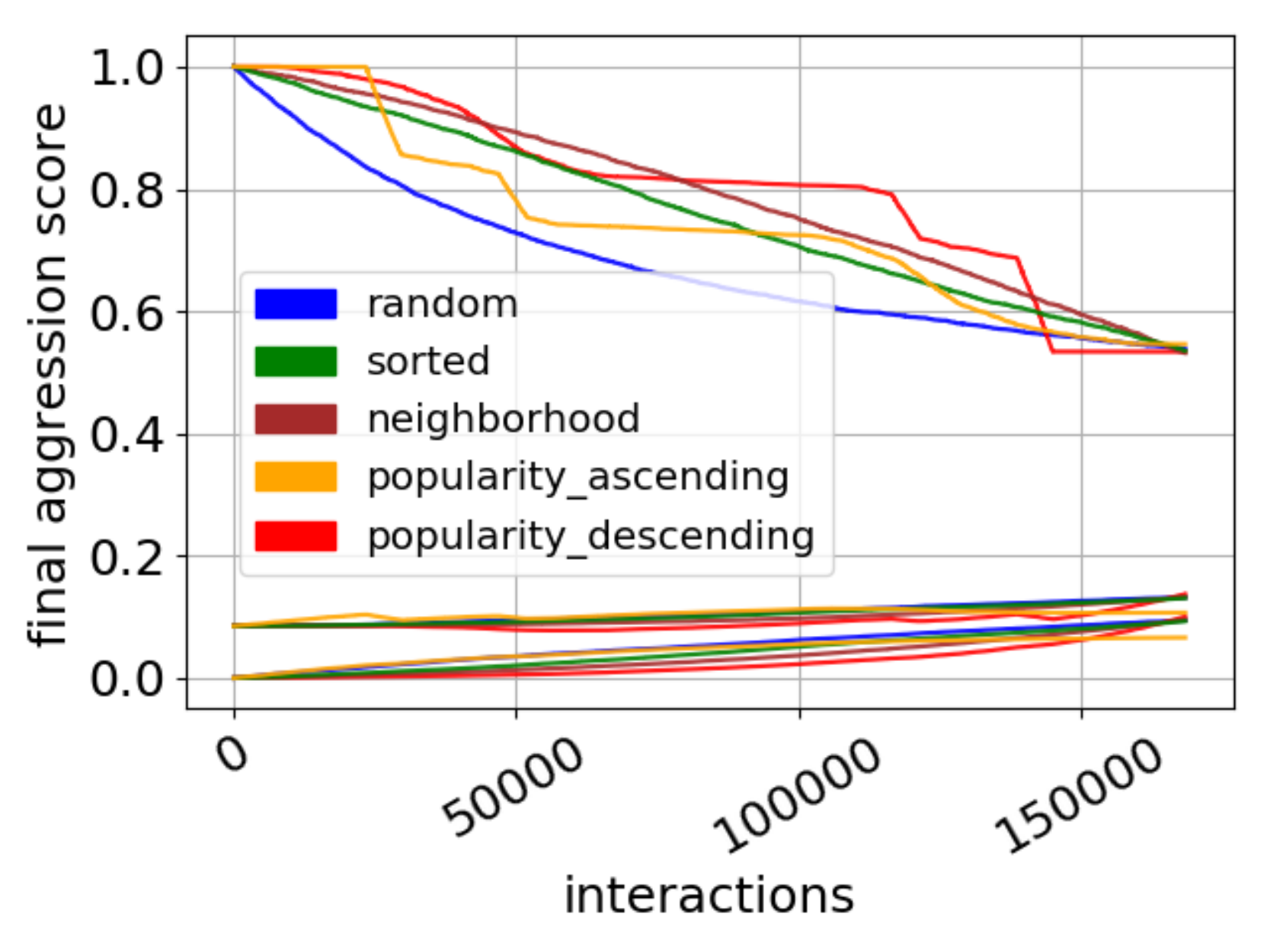}} 
    \subfigure[HK\_0.5\_P]{\includegraphics[width=8cm]{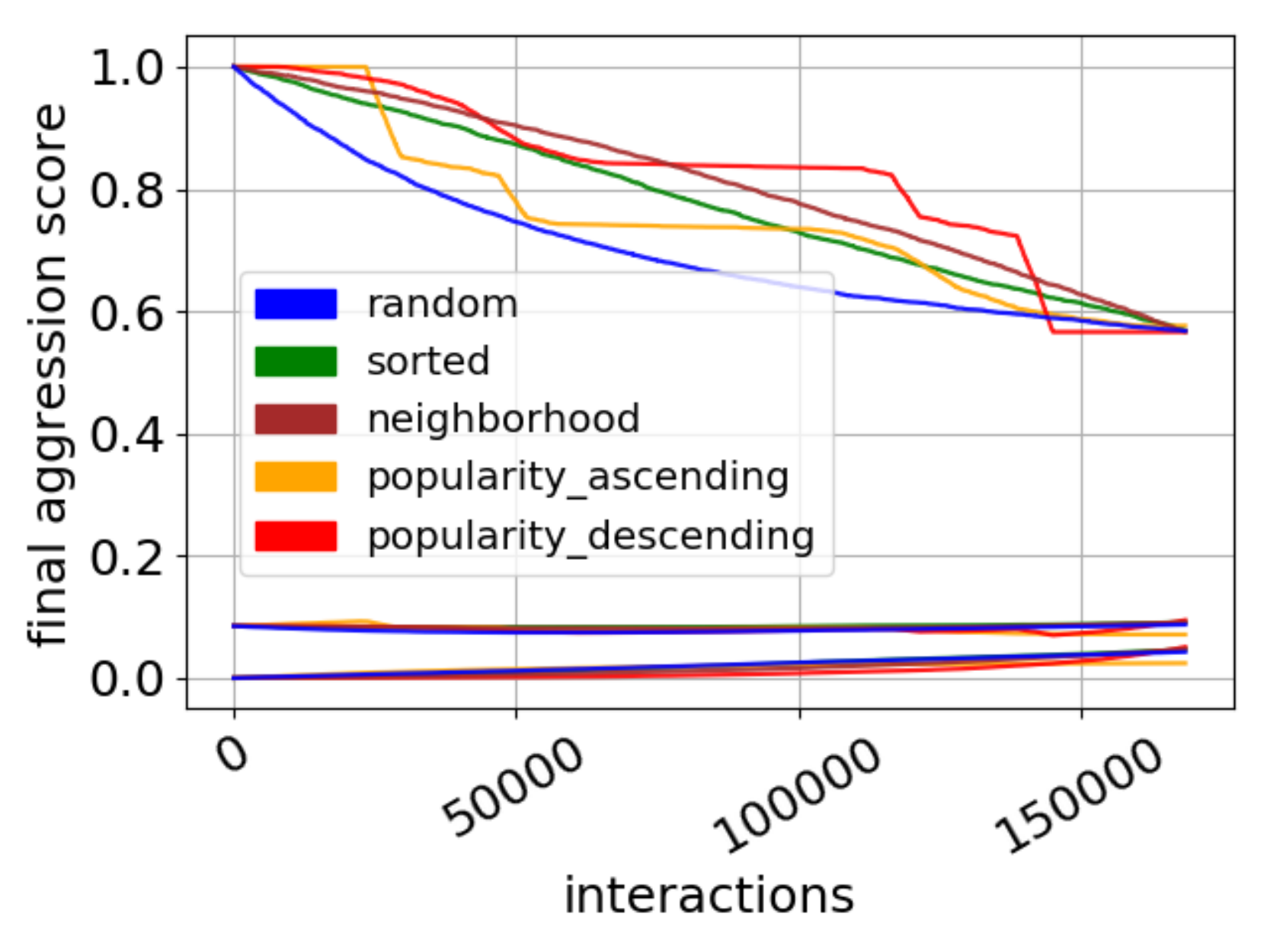}}
    \subfigure[HK\_1.0\_P]{\includegraphics[width=8cm]{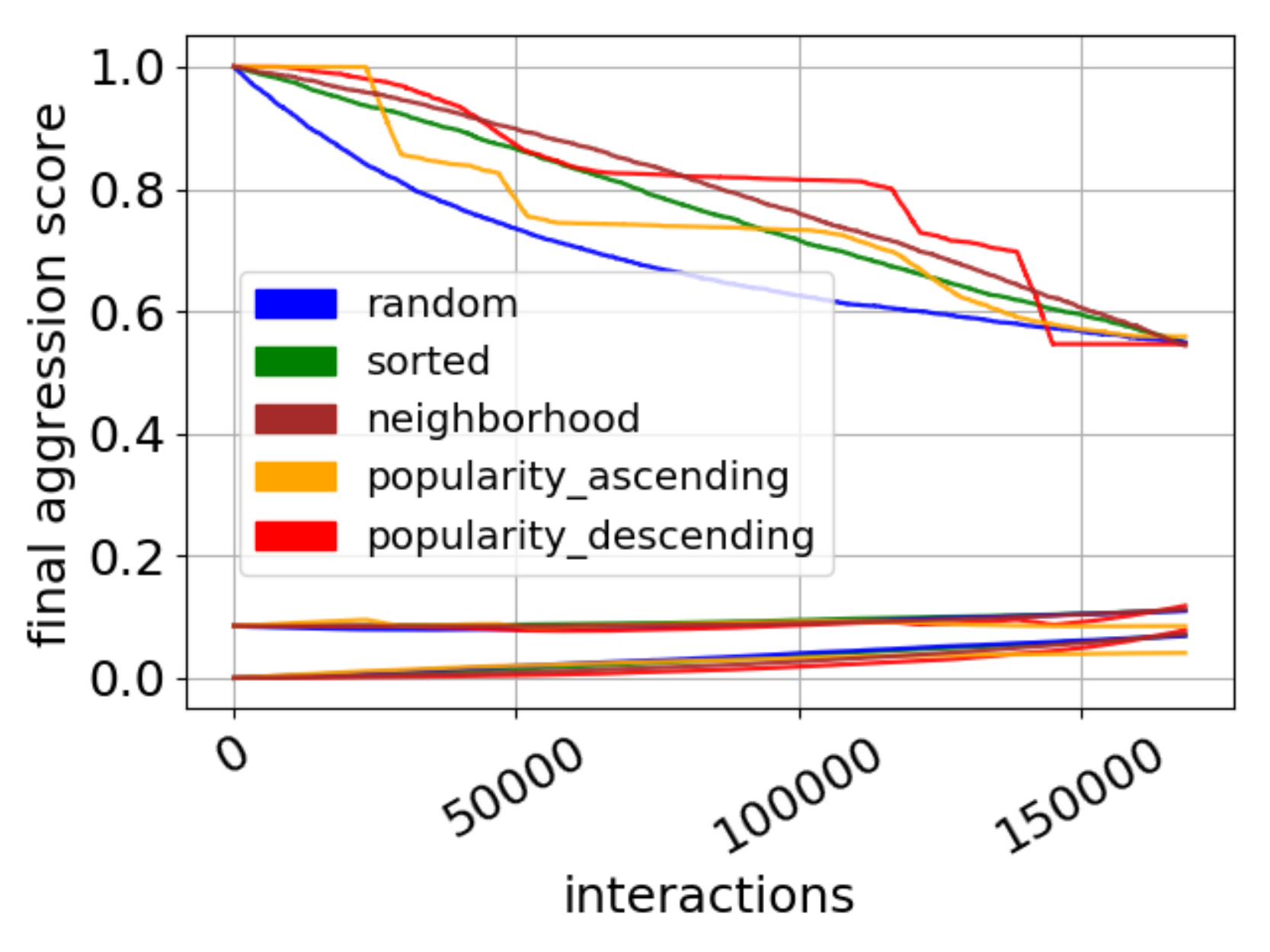}}
    \caption{Final average aggression score for aggressive (top part), normal (bottom part), and all (middle part) users, based on 5 different types of users' ordering, through the simulation time, for the top four models.
    }
    \label{fig:finalAggressionScoreOrder}
\end{figure*}

Figure~\ref{fig:finalAggressionScoreOrder} shows the aggression evolution of three sets of users (i.e., normal, aggressive, all users) in relation to how they could interact (i.e., randomly, based on popularity (most/least), involved neighborhood, and network id), for the top four models.
Specifically, the top and bottom parts of the plots depict how the aggression changes for the aggressive and normal users, respectively, while the middle includes all users.
If aggressive users were to interact randomly, it would lead to faster decline of the aggression compared to the rest of the ordering methods.
In contrast, aggressive users show greater resistance in reducing their aggression if they were to interact based on users' popularity (from highest to lower).
For instance, authors in~\cite{corcoran2015cyberbullying} observed that more powerful (i.e., popular) users have a stronger impact on others' feelings.
Interestingly, assuming the least popular users were to interact (act on their aggression) first, it would lead to slower propagation.
If aggression propagated from one neighborhood to the next, it would also lead to a slow rate of propagation, at times approaching the least popular user ordering.
For normal users, aggression status could not be significantly affected by the way they interact; the difference between initial and final aggression scores is subtle.
As we see in Figure~\ref{fig:finalAggressionScoreOrder}, the difference between the initial and final aggression scores of normal users is subtle, indicating that normal users are particularly resistant in the aggression that may exist within their network.
Instead, the effect of normal users is greater on aggressors than the inverse.
For all considered models (apart from \textit{Voter}) after a number of interactions (i.e., $175K$), the aggression score converges, indicating that within a network, even with faster or slower changes, the network state stabilizes.

\descr{Takeaways.}
Overall, the way users could interact and ``exchange'' or propagate aggression impacts the overall network state.
Popularity of users can be a great predictor of how aggression will move in the network.
This could be attributed to the fact that more popular users can have stronger impact within a network if they are aggressive (or not) due to their high degree centrality, since they can affect many users at the same time, leading to high rate of aggression propagation.
This is also aligned with phenomena already evident in the wild (3rd insight: \textit{Influence due to power difference}, Sec.~\ref{sec:insightPowerDifference}).
At the same time, normal users are more resistant to aggression, due to their expected higher power status and larger neighborhoods with non-aggressive state.

\begin{figure*}[!t]
\center
	\subfigure[Normal users]{\includegraphics[width=7cm]{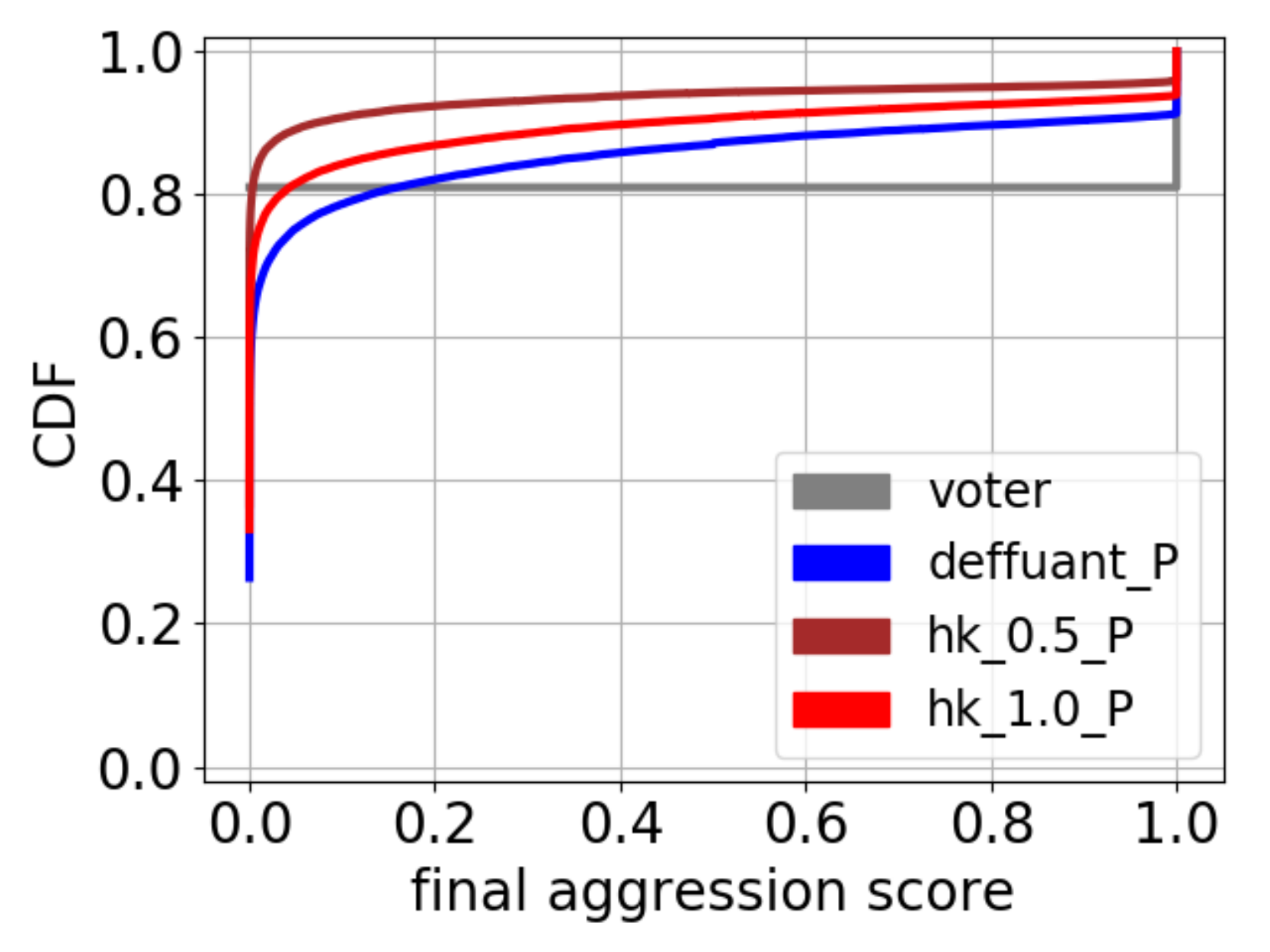}
	\label{fig:nn_cdfFinalAggressionScoreNormalUsers}}
	\subfigure[Aggressive users]{\includegraphics[width=7cm]{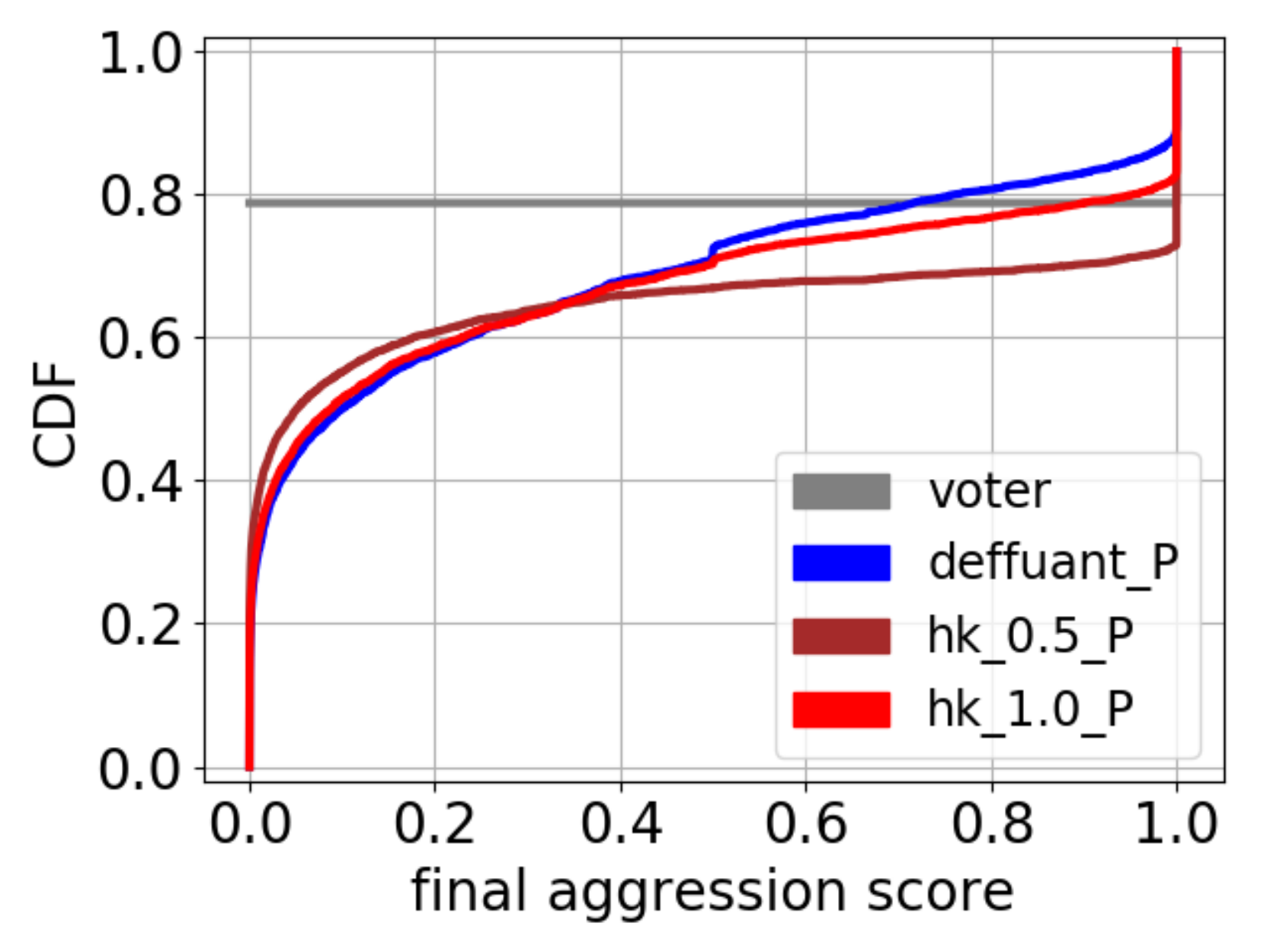}
	\label{fig:nn_cdfFinalAggressionScoreAggressiveUsers}} 
	\caption{CDFs of the final aggression scores of (a) normal and (b) aggressive users who participated in the aggression propagation.}
	\label{fig:nn_cdfFinalAggressionScoreAggressiveNormalUsers}
\end{figure*}

\subsection{What is users' final aggression state?}\label{sec:final-aggression-score}

Figure~\ref{fig:nn_cdfFinalAggressionScoreAggressiveNormalUsers} shows how users' aggression has settled at the end of the simulation (on 10th snapshot) across the top four performing models; we consider the more realistic random ordering of changes.
Figure~\ref{fig:nn_cdfFinalAggressionScoreNormalUsers} shows that regardless of model, at least $60\%$ of normal users in the end remain unaffected (i.e., their aggression score is zero), while at most $40\%$ of users gain some aggression (at different levels, varying from $0$ to $1$).
For instance, based on \textit{Deffuant\_P}, almost $10\%$ of users end up with maximum aggression score, with the rest of users varying between the lower and upper limits.
In \textit{Voter}, because of its formulation, the aggression score is either $0$ or $1$, with about $20\%$ of normal users affected.
Finally, from Figure~\ref{fig:nn_cdfFinalAggressionScoreAggressiveUsers}, about $20\%$-$30\%$ of aggressive users maintain their high aggression score (depending on the model), with only about $30\%$ of aggressive users turning normal (\textit{Deffuant\_P} and \textit{HK\_*\_P}).
Contrary to normal users, when $Voter$ is used, a high portion of $\sim80\%$ aggressive users is positively affected by turning into normal.
Overall, in consistency with Figure~\ref{fig:finalAggressionScoreOrder}, normal users are more resistant to adopting aggression, as opposed to aggressive users.

\subsection{Validation of top models in prediction}
We conclude this investigation by evaluating the performance of the top four models in predicting the aggression status of users,
by testing the models on the ground truth dataset \textit{Crawled Net} $LT$ (Table~\ref{table:datasets}).
However, this network is comprised of the $401$ users' ego-networks, i.e., their friends and followers, but not their 2nd-hop and beyond (we could not collect this information due to Twitter limitations).
To perform fair application of the models, we filtered this graph and kept a subgraph that includes all ego users (out of the $401$) that have edges connecting them.
This way, all users participating in the experiment have their ego-graphs crawled (i.e., $i$ and $j$ in the models).
As shown in Table~\ref{table:datasets}, this filtering resulted in a subgraph of $314$ users and $540$ edges (\textbf{Ego-Ego Net} $LT$).
Then, we kept the graph's SCC and concluded with $57$ nodes and $395$ edges.

Due to its small size, we select all the edges of this graph for the interaction (and not only a random 10\% of them).
This simulation tests all the different types of users' ordering (Sec.~\ref{sec:orderingofchanges}) and the four best models:
\textit{Voter}, \textit{Deffuant\_P}, \textit{HK\_0.50\_P}, and \textit{HK\_1.0\_P}.
Different aggression thresholds $T_A=0.05,~\dots,~0.9$ are used to characterize a user as aggressive if their final state is greater than the threshold, and normal otherwise.
The goal of this experiment is to let the models simulate propagation of aggression, thus altering users' state.
Then, we measure their performance in predicting correctly the final state of users (aggressive or normal, i.e. true positives or negatives, respectively), as found in \textbf{Ego-Ego Net} $LT$, using standard ML metrics such as Precision, Recall and AUC (using different operating $T_A$s).

\begin{table*}[t]
	\centering
	\caption{Model performance at predicting which users will remain (or change into) normal or aggressive.
	(Showing only the results for AUC $\geq$ 0.70, on the \textit{Ego-Ego Net} SCC $LT$).}
	\label{table:auc-small-network-ego-ego}
\scalebox{1.20}{
	\begin{tabular}{l l l l l l l}
		\toprule
		\textbf{Model} & \textit{Order} & \textit{$T_A$} & \textit{\#edges} & AUC & Precision & Recall\\
		\midrule
		Deffuant\_P & sorted & 0.10 & 395 & 0.71 & 0.59 & 0.71\\
		Deffuant\_P & sorted & 0.20 & 395 & 0.74 & 0.60 & 0.74\\
		HK\_1.0\_P & sorted & 0.10 & 395 & 0.72 & 0.60 & 0.72\\
		HK\_1.0\_P & sorted & 0.20 & 395 & 0.75 & 0.60 & 0.75\\
		HK\_1.0\_P & random & $[0.05, 0.30]$ & 50 & 0.76 & 0.76 & 0.76\\
		Deffuant\_P & random & $[0.05, 0.30]$ & 50 & 0.79 & 0.70 & 0.79\\
		\bottomrule
	\end{tabular}
	}
\end{table*}

From Table~\ref{table:auc-small-network-ego-ego}, we observe that the best performance arises when the interaction of users happens based on the increasing ranking of their user IDs, indicating that account age (which is proxied by the user ID) may have important influence in the propagation process.
We notice that \textit{Deffuant\_P} and \textit{HK\_1.0\_P} achieve good performance when low $T_A$ is used, such as $0.10$-$0.20$, and the AUC score ranges from $0.71$ to $0.75$.
In addition, we experimented with executing the models for a portion of the edges available.
Using random users' ordering, the \textit{HK\_1.0\_P} model achieves AUC score, Precision and Recall equal to $0.76$, for $T_A$=$[0.05, 0.30]$, after $50$ interactions.
Also, satisfactory validity was achieved for \textit{Deffuant\_P} method, when 50, 100, 200 and 250 interactions take place, in random order.  
We observe that after $50$ interactions, and while $T_A$ ranges from $0.05$ to $0.30$, the AUC score, Precision and Recall are equal to $0.79$, $0.70$ and $0.79$, respectively.

\descr{Takeaways.}
Overall, with these validation experiments, we show that our top performing models are capable of emulating reliably the propagation of aggression, as validated on this small, annotated Twitter dataset.
In addition to the previous results, we find that older accounts propagating first their aggression to newer accounts matches better the ground truth data at hand.
This hints that account age may have important influence in the aggression propagation process and renders some extended future work.

\section{Conclusion \& Future Work}

Despite the consequences that aggressive behavior has on individuals (e.g., embarrassment, depression, isolation from other community members, etc.), there are still important cases of aggression that stay under the radar of social networks, e.g., as in~\cite{cesarbomb}.
In fact, how such behavior propagates within networks and which network or other factors affect this propagation, have not been extensively studied.
To address this gap, here, we are the first to propose a pipeline to evaluate various aggression dynamics models, and to conclude in those best emulating aggression propagation in social networks.
To simulate how such behavior could spread in the network, we built on top of popular opinion dynamics models, and test and validate our models' performance on real Twitter data.
We found that our proposed models based on the \textit{Deffuant}  and \textit{Hegselmann \& Krause} opinion models, perform best in modeling aggression propagation in a network such as Twitter, regardless of parameters or thresholds used.
Important insights embedded in these models are:
(1) online aggression tends to propagate from one user to another, (2) power score of a user (e.g., degree centrality) and (3) users' internal aggression state, both constitute top factors to be considered in aggression propagation modeling,
(4) influence by users' neighborhood is of less importance than other factors.

Overall, we believe this work makes a significant first step towards understanding and modeling the dynamics of aggression.
The outcomes of our work highlight the suitability of the top performing models in simulating propagation of aggression in a network such as Twitter, and how a campaign to monitor and even stop aggression on Twitter could work.
That is, if aggressive users are monitored in their interactions with others (e.g., posting of aggressive messages), and simultaneously, normal users are shielded from this aggression by dropping such communication, the overall aggression in the network will significantly drop.
In fact, if the campaign targets highly popular aggressive users, who are encouraged to reduce their aggression via educational tutorials and other interventions, the overall aggression in the network can drop faster than selecting users with different criteria (e.g., random).
Also, we found evidence that the age of an account may be influencing the way aggression propagates.
However, more in-depth investigation on ground truth data with aggression propagation information is needed to reveal clearer patterns of this aspect.

As first proposed by~\cite{squicciarini2015bully-pairwise}, and moving towards a better, more global or unified model of aggression propagation, in the future, we can consider aspects of this behavior in relation to social media activity, such as the actual content of a post, the time it was posted, which user posted it, which users commented on the post and how long after the original post was done, the reaction of the neighborhood of the original poster and commenters, etc.
Moreover, Linear Threshold and Independent Cascades are two alternative sets of models that could be examined and compared with the opinion dynamics based models for modeling aggression propagation.
Early work on these models has recently showed promising results~\cite{poiitis2021aggressionpropagationICLT}, but more in-depth analysis of such models and on different datasets is needed.
Another interesting extension of this work would be to attempt aggression propagation modeling on a dynamic network.
Evolving networks, where links are added or removed through time, are the most realistic but notoriously difficult to model, and render significant future attention.
Finally, it would be worthwhile to investigate the effectiveness of the proposed models to predict aggression on other online social network platforms such as Facebook, Instagram, etc.

\begin{acks}
This research work has been partially funded by the European Union’s Horizon 2020 Research and Innovation program under the Marie Skodowska-Curie ENCASE project (Grant Agreement No. 691025) and CONCORDIA project (Grant Agreement No. 830927).
Also, it has been supported by the Hellenic Foundation for Research and Innovation (H.F.R.I.) under the ``1st Call for H.F.R.I. Research Projects to Support Faculty Members \& Researchers and Procure High-Value Research Equipment'' (Project Number: HFRI-FM17-1873, GraphTempo), and the Erasmus+ Programme, Higher Education 2018-19, KA1 Learning Mobility of Individuals (Grant Agreement 2018-1-EL01-KA103-046833).
\end{acks}

\bibliographystyle{plain}
\bibliography{main}

\end{document}